\documentclass[preprint,12pt ]{elsarticle}

\usepackage[margin=1in,a4paper]{geometry} 

\usepackage{amssymb}
\usepackage{amsmath}
\usepackage{bm,bbm}
\usepackage{tabularx} 
\usepackage{graphicx} 
\usepackage{float}
\usepackage{caption}
\usepackage{subcaption}
\usepackage{comment}
\usepackage[final]{hyperref} 
\hypersetup{
	colorlinks=true,       
	linkcolor=blue,        
	citecolor=magenta,        
	filecolor=cyan,     
	urlcolor=cyan         
}

\usepackage{cleveref}

\crefformat{equation}{(#2#1#3)}
\crefrangeformat{equation}{(#3#1#4) to~(#5#2#6)}
\crefrangeformat{equation}{(#3#1#4) to~(#5#2#6)}

\journal{Journal of Computational Physics}

\begin{document}

\begin{frontmatter}



\title{Improving the performance of regularized Stokeslet simulations using Rosenbrock time integration methods} 


\author[L1]{Moslem Uddin\corref{cor1}} 
\cortext[cor1]{Corresponding author}
\ead{muddin1@tulane.edu}
\author[L1]{ Lisa J. Fauci   } 
\author[L1]{  Ricardo Cortez } 
\author[L1]{ Tommaso Buvoli } 
\affiliation[L1]{organization={Dept. of Math, Tulane University},
            addressline={6823 St. Charles Av.}, 
            city={New Orleans},
            postcode={70118}, 
            state={LA},
            country={USA}}
\begin{abstract}

The method of regularized Stokeslets (MRS) is widely used to model fluid-structure interactions of microscopic bodies in viscous environments. 
Specifically, MRS computes the fluid velocity generated by regularized forces distributed on an immersed body, which in turn defines a system of ordinary differential equations with which to evolve the body so that the no-slip condition is enforced.  
To accurately model material properties such as elasticity and bending rigidity, the resulting ODE system is often stiff and,  therefore,  computationally expensive to solve with explicit time-integration methods. Here we demonstrate the potential of linearly implicit Rosenbrock (ROS) time integration methods to greatly reduce the computational time required to solve such ODE systems. We focus on the classic model problem of the swimming of a flexible filament driven by preferred curvature forcing. Our findings indicate that, compared to previous MRS simulations, the increased stability of ROS methods allows for speed-ups between twenty and two thousand times over explicit Runge-Kutta (ERK) methods.
Moreover, we present approximate scaling laws for this speedup as a function of the key model parameters (bending rigidity and slenderness) at the root of numerical stiffness.  
Lastly, beyond the basic ROS approach, we discuss several additional strategies that further improve performance.
\end{abstract}






\end{frontmatter}




   \section{Introduction}\label{s1}

 The study of biological fluid 
 dynamics is fundamental to understanding a wide range of physiological and biological processes, from blood flow in arteries to the locomotion of microorganisms. These coupled systems give rise to fluid-structure interactions where the immersed structure (e.g. blood vessel, cell membrane, flagella) imparts forces to the surrounding fluid, but, at the same time, moves in response to the evolving flow.  
At the microscale, where viscous forces dominate, inertia may be neglected and the fluid flow is characterized by the Stokes equations. 
Because the incompressible Stokes equations are linear, they admit fundamental solutions, and these have been leveraged in some widely used numerical methods 
to study  FSI problems such as slender body theory \citep{Keller_Rubinow_1976,Lighthill76, Nguyen01112014}, resistive force theory \citep{Gray55}, the boundary element method \citep{Pozrikidis_1992, ISHIKAWA_SIMMONDS_PEDLEY_2006}, the method of regularized Stokeslet (MRS) \citep{Cortez01, Cortez05}, and the immersed boundary method \citep{BRINGLEY20085397, Teran09}. 
 
 This article focuses on time-stepping within the method of regularized Stokeslets (MRS). In each time step of the coupled fluid-structure system, the MRS evaluates the resulting velocity at all discrete points on the immersed structure due to regularized forces concentrated at these discrete points. 
The motion of the immersed structure is governed by the no-slip condition, which says that the velocity of these discrete points must be the same as the fluid velocity evaluated at those points. 
Therefore, if we denote the position of the structure as $\bm{x}$, this  leads to a system of ordinary differential equations (ODEs):
 \begin{align}\label{e11}      	 	
 	\dfrac{d \bm{x} }{dt}&=\bm{u}(t,\bm{x} ),\quad  \bm{x}( 0) = \bm{x}_{0}.
 \end{align}
 
 In general, $\bm{u}$ is a nonlinear function of structure geometry and material parameters that leads to a stiff system, which is computationally expensive to evolve \citep{BOUZARTH20104208}.  Despite this, explicit Runge-Kutta (ERK) methods have been widely used in conjunction with MRS, leading to the requirement of a very restricted step size for stable solutions
\citep {OLSON2011203, SIMONS201481}. For instance, in the investigation of the propulsion of a flexible undulating flagellum  \citep{OLSON2011203} a stable solution using a fourth-order ERK method required  $2\times 10^6$ 
time steps per flagellar beat period.  In this study related to the locomotion of hyperactivated mammalian sperm, physiological parameters associated with bending rigidity of mice sperm were used.  Given that a typical mice sperm requires on the order of 20 flagellar beats to swim one body length, simulations that track multiple, or even a single model flagellum, require a prohibitive amount of time steps using ERK methods. 

In this article, we revisit the model problem of a single flexible flagellum driven by preferred curvature forcing.
 The step size restriction for ERK methods motivated us to consider linearly implicit Rosenbrock (ROS) integrators \citep{Rosenbrock63, Wanner1977,wanner1996}, which require linear solver at each time step \citep{wanner1996}. ROS methods exhibit admirable stability properties that allow for larger step sizes and a reduction in computational time.
 Various other time integrators have been previously proposed for accelerating biofluid simulations. This includes implicit methods \citep { SCHOELLER2021, Townsend_Keaveny_2024}, spectral deferred correction methods, \citep{MALHOTRA2024}, multirate methods \citep{BOUZARTH20104208, Nourian_Shum_2023}, and time-parallel methods \citep{LIU2022111366}.
  However, to the best of our knowledge, ROS methods have not been used to enhance the computational performance of MRS biofluid simulations.

 We have organized this article as follows: in sections \ref{s2} and \ref{s3}, we review the MRS and time-stepping methods. In Section \ref{s4}, we describe the biofluid example that is investigated in this study. Section \ref{s5} will outline the causes of stability restrictions for ERK methods, based on model parameters. Moreover, we will provide scaling laws for the speedup achievable with ROS methods for the problem considered in this section.  Furthermore, we will discuss the performance gains observed with the ROS method, along with directions for further improvement in Section \ref{s6}. Lastly, we present concluding remarks and future directions in Section \ref{s7}.  
   
   \section{Review of Method of Regularized Stokeslets}\label{s2}

The steady Stokes' equations together for a viscous, incompressible flow in $\mathbb{R}^3$ are: 
\begin{align}
	\label{e1} \mu \Delta \bm{u}&=\nabla p-\bm{\Tilde{F}},\\      
	\nabla \cdot \bm{u}&=0,\label{e2}
	 \end{align} 
	  where $\mu$ is the viscosity of the fluid, $p$ is the pressure, $\bm{u}$ is the velocity, and $\bm{\Tilde{F}}$ is the external force per unit volume applied to the fluid.
	The method of regularized Stokeslets considers the regularized force density $\bm{\Tilde{F}}= \bm{\tilde{f} \phi _ {\epsilon }(\bm{x}-\bm{\Tilde{x}}} )$ concentrated at the point $\bm{\Tilde{x}}$ \citep{Cortez05, CORTEZ201541}.  For each choice of the  parameter $\epsilon >0$, the chosen regularization function (``blob'' function) is radially symmetric, smooth, and has the property: 
	 $\int_{\mathbb{R}^3} \phi_{\epsilon}(\bm x)d\bm{x}=1.$
     
	  Here we use the standard MRS 
       blob function:  
       \begin{eqnarray}
       \phi_\epsilon(r)=\dfrac{15\epsilon^4}{8\pi(r^2+\epsilon^2)^{\frac{7}{2}}},  
       \end{eqnarray}
       where $r=|\bm x-\bm {\Tilde{x}}|$.
       With this choice, the velocity at any point $\bm{x}$ can be computed analytically to give the regularized Stokeslet solution: 
	   \begin{align}
	   	    \bm{u}(\bm x)=&\bm{\Tilde{f}}\dfrac{r^2+2\epsilon^2}{8\pi\mu(r^2+\epsilon^2)^{\frac{3}{2}}}+\dfrac{(\bm{\Tilde{f}} \cdot(\bm x-\bm{\Tilde{x}}))(\bm x-\bm{\Tilde{x}})}{8\pi\mu(r^2+\epsilon^2)^{\frac{3}{2}}}=\bm S(\bm x,\bm {\Tilde{x}}) \bm{\Tilde{f}}.\label{RS3D}
	   	    \end{align}
	 
As the Stokes equation are linear, the velocity field due to forces concentrated at $N$ discrete nodes of the immersed object is the superposition of the $N$ regularized Stokeslets. 
Imposing the no-slip boundary condition at each of the $N$ discrete nodes of the object therefore leads to the ODE system:
\begin{eqnarray}
     \frac{d\bm{\Tilde{x}}_j}{dt}=\bm{u (\bm{\Tilde{x}}_j)}=\sum_{k=1}^N \bm{S}(\bm{\Tilde{x}}_j,\bm{\Tilde{x}}_k) \bm{\Tilde{f}}_k
     \label{systemode}
\end{eqnarray}

\noindent
Here $\bm{S}$ is the regularized Stokeslet kernel in Eqn. \ref{RS3D} above, and the model-dependent forces $\bm{\Tilde{f}}_k$ typically represent passive elastic forces as well as time-dependent actuation forces supported on the immersed structure. The time-stepping of the ODE system in Eqn. \ref{systemode} is the focus of this work.

\section{Review of Time Integration Methods} \label{s3}
This section reviews the time integration strategies that will be considered in this article, namely explicit Runge-Kutta and Rosenbrock methods. For brevity, we present all schemes by assuming an autonomous ODE of the form $\bm{x}'=\bm{u}(\bm{x})$; the corresponding non-autonomous form is provided in \ref{app1}.
\subsection{Runge-Kutta (RK) Methods}\label{s31}
   RK methods  \citep{butcherch3} are widely used to solve ODE systems arising from various physical systems.   An $s$-stage RK method is:
   \begin{subequations}\label{e31}
\begin{align}
	\bm{K_i}&= \Delta t\bm{u}( \bm{x}_n+ \sum_{j=1}^{s} a_{ij} \bm{K_i}),\quad i=1,\ldots,s\label{e32}\\
	\bm{x}_{n+1}&= \bm{x}_n+ \sum_{i=1}^{s} b_i \bm{K_i}, 
\end{align}
\end{subequations}
 where $\Delta t$ is the time step, $\bm{x}_n$ denotes the approximate solution at $t =t_n$, and the coefficients $(a_{ij}, b_i )$ define a specific RK method. The sparsity of the matrix $a_{ij}$ determines the type of the method.
For example, explicit Runge-Kutta (ERK) methods are characterized by a strictly lower triangular $ a_{ij}$ \citep{hairer1993solving}.

In this study, we will consider two well-known ERK methods: forward Euler \citep{butcherch3} and a third-order Heun ERK \citep{hairer1993solving}, which we refer to as RK1 and RK3, respectively.
\begin{align}\label{erk}
  \text{\bf RK1:} \quad 
  \begin{aligned}
    \bm{K_1}&= \Delta t\bm{u}( \bm{x}_n ), \\
    \bm{x}_{n+1}&= \bm{x}_n+ \bm{K_1},
  \end{aligned}
  \qquad \qquad \qquad
  \text{\bf RK3:} \quad 
  \begin{aligned}
    \bm{K_1}&= \Delta t\bm{u}( \bm{x}_n ), \\
    \bm{K_2}&= \Delta t\bm{u}( \bm{x}_n +\dfrac{1}{3} \bm{K_1}), \\
    \bm{K_3}&= \Delta t\bm{u}( \bm{x}_n  +\dfrac{2 }{3} \bm{K_2}),   \\
    \bm{x}_{n+1} &= \bm{x}_n+\dfrac{1}{4} (\bm{K_1}+3\bm{K_3}).
  \end{aligned}
\end{align}

The primary motivation for choosing RK1 is its widespread use in MRS-based biofluid problems (e.g.\cite{yijiang-graham, CORTEZ2018,JABBARZADEH2020,ferranti2025,Morshed25}. On the other hand, we selected RK3 after finding that it provides a better balance between accuracy and computational efficiency compared to second and fourth order ERK methods.
\subsection{Rosenbrock Methods}\label{s32}

Rosenbrock (ROS) methods only require linear solves at each stage, as opposed to the nonlinear solves required by implicit Runge-Kutta methods. An $s$-stage 
ROS method \citep{Rosenbrock63,Wanner1977,wanner1996} is:
\begin{subequations}\label{eq8}
\begin{align}
		\bigg(\bm{I}-\gamma_{ii}\Delta t\bm{J}_n
	  \bigg)\bm{K_i} &= \Delta t\bigg( \bm{u}( \bm{x}_n +
	\sum_{j=1}^{i-1} \alpha_{ij}\bm{K_j} ) +  \bm{J}_n
	\sum_{j=1}^{i-1}\gamma_{ij} \bm{K_j} \bigg)   ,\quad i = 1, \ldots , s,\label{eqn_row}\\
	\bm{x}_{n+1} &= \bm{x}_n +
	\sum_{i=1}^{s} b_i \bm{K_i},
\end{align}
\end{subequations}
where the coefficients $(\alpha_{ij}, \gamma_{ij}, b_i)$  define a specific ROS method.  Moreover, $\bm{J}_n = \frac{\partial \bm{u}}{\partial \bm{x}}( \bm{x}_n)$ represents the Jacobian of $\bm{u}$ evaluated at $ \bm{x}_n$ and $\bm{I}$ is the identity matrix. One of the simplest ROS methods is first-order ROS-Euler (which we referred to as ROS1) \citep{wanner1996}:
\begin{align}\label{ROS1}
\begin{aligned}
       \bigg(\bm{I}- \Delta t\bm{J}_n
	  \bigg) \bm{K_1} &= \Delta t \bm{u}( \bm{x}_n ) , \\
	\bm{x}_{n+1} &= \bm{x}_n + \bm{K_1}.
    \end{aligned}
\end{align}
In practice, ROS methods \eqref{eq8} are typically implemented using an alternative formulation that avoids the matrix-vector multiplications $ \bm{J}_n
	 \bm{K_j}$ \citep{wanner1996, Lang2021}. These transformations are discussed in \ref{app1}. Moreover, the implicit coefficients often satisfy $\gamma_{ii}=\gamma,  i=1,\ldots,s$ \citep{Lang2021}; for constant timestepping, this allows one to perform a matrix factorization of $\bm{I}-\gamma\Delta t \bm{J}_n$ to compute all linear solves more efficiently.
 
All ROS methods require solving $s$ linear systems involving the Jacobian of the ODE right-hand side. There are several approaches for computing the Jacobian matrix, with an analytical derivation being the most straightforward. However, in many cases, this is either intractable or exceedingly complicated due to the complexity of the underlying equations. In the following subsections, we discuss two well-known alternatives.

\subsubsection{Computing the Jacobian -- Finite Differences}
 A widely used method for computing the Jacobian is the first-order (forward) finite difference (FD) approximation. This approach estimates the $i$th column of the Jacobian matrix using
\begin{align} 
\frac{\partial \bm{u}}{\partial x_i}(\bm{x}) = \frac{\bm{u}(\bm{x}+h \bm{e}_i )-\bm{u}(\bm{x})}{h}+\mathcal{O}(h), \quad i=1,\ldots,N,\label{e47}
\end{align}
where $\bm{e}_i$ denotes the usual basis vector, $h$ is a small perturbation parameter, and $N$ represents the spatial dimension of the system. Constructing the Jacobian using this approach requires $N+1$ evaluations of  $\bm{u}(\cdot)$.  

While the FD approach is conceptually simple to implement, it only provides an approximation to the Jacobian and is susceptible to truncation and rounding errors \citep{Atilim18}. This can be problematic because a ROS method of order $p$ requires that the Jacobian approximation be accurate to order $p$. However, achieving this accuracy with first-order finite-difference (FD) approximations can be challenging due to round-off errors as $h\to 0$. The next approach we discuss can circumvent this issue.

\subsubsection{Computing the Jacobian -- Automatic Differentiation}
Recent advances in scientific machine learning offer a promising alternative to traditional finite-difference approaches by leveraging automatic differentiation (AD) to construct Jacobians. Unlike FD approaches, AD computes derivatives to machine precision by systematically applying symbolic differentiation rules to elementary functions. As a result, AD eliminates the truncation errors that are inherent to FD approximations \citep{Atilim18}.  In subsection \ref{ss73} we will demonstrate this on a biofluid problem.
\subsection{W-methods}\label{ss33}

Recomputing the Jacobian at every time step can introduce a major computational cost for ROS methods. $W$-methods are designed to address these limitations by allowing for an inexact Jacobian. There are two common flavors of $W$-methods: those that retain order-of-accuracy when the Jacobian is replaced with an {\em arbitrary matrix}, and those that require an $\mathcal{O}(\Delta t)$ approximation to the Jacobian. We refer to the first as ROW methods and the second as $RO\widetilde{W}$ methods.

\subsubsection{Classical $W$-methods (ROW)}
The family of $W$-methods is structurally identical to ROS methods, except that the Jacobian can be replaced with an arbitrary matrix $\bm{W}_n$ \citep{wanner1996, Lang2021}. 
A simple example of a $W$-method is the $W$-Euler method:
\begin{align}\label{ROW1}
\text{\bf ROW1:} \quad
\begin{aligned}
       \bigg(\bm{I}- \Delta t\bm{W}_n
	  \bigg) \bm{K_1} &= \Delta t \bm{u}( \bm{x}_n ) , \\
	\bm{x}_{n+1} &= \bm{x}_n + \bm{K_1}.
    \end{aligned}
\end{align}
We note that this method is identical to the ROS-Euler method \eqref{ROS1}, except for $\bm{J}_n \to \bm{W}_n$.

\subsubsection{$\widetilde{W}$-methods ($RO\widetilde{W}$)}
There are also variants of $W$-methods that require more accurate  Jacobian approximations  \citep{verwer1980rosenbrock,jax2020rooted}. Specifically, at the $n$th time step $\bm{W}_n$ must be an order $\mathcal{O}(\Delta t)$ approximation to the exact Jacobian (\textit{i.e.} $\bm{W}_n=\bm{J}_n + \mathcal{O}(\Delta t)$); we will refer to such integrators as $\widetilde{W}$-methods. Although $\widetilde{W}$-methods are less flexible than classical $W$-methods, $\widetilde{W}$-methods require fewer order conditions to derive while still proving more flexibility than ROS methods.

\subsection{ERK, ROS, and W-methods considered in this work}

In table \ref{tabm}, we list the ERK, ROS, $W$,  and $\widetilde{W}$-methods that we will consider in the article along with their properties. 

\begin{table}[H]
    \centering
    \begin{tabular}{|l|c|c|c|c|l|}
    \hline  Label& Coefficients& Order & Stages &Type & Reference \\\hline 
           RK1 &\ref{erk} & 1&1 &ERK    &\citep{butcherch3}\ \\
       ROS1 &\ref{ROS1} & 1&1 &ROS    &\citep{wanner1996}\ \\
       ROW1 &\ref{ROW1}& 1&1 &$W$    &\citep{wanner1996}\ \\
              RK3 &\ref{erk}& 3&3 &ERK    &\citep{hairer1993solving}\ \\
      ROS3 &\ref{ROS3}   & 3 &3& ROS    & \citep{SANDU19973459}\\
      ROW3 &\ref{ROS34PW1a}   & 3&4 &   $W$   & ROS34PRW1a from \citep{rang_new_2005}\\
           \text{RO$\widetilde{W}$3} &\ref{GROW37n2}  & 3&7 &  $\widetilde{W}$   & GROW37n2 from \citep{jax2020rooted}\\
\hline
    \end{tabular}
    \caption{ERK, ROS, $W$, and $\widetilde{W}$-methods considered in this article.}
    \label{tabm}
\end{table}

 We selected ROS1 and ROW1, the ROS/$W$  counterparts of RK1, due to their straightforward formulations and superior stability property among all ROS, $W$, and $\widetilde{W}$ methods we tested from \citep{Lang01,Atilim18,RANG2015128,ROS2,wanner1996}. Similarly, we selected ROS3, ROW3,  and \text{RO$\widetilde{W}$3}, because they offered the best balance between stability, accuracy, and computational cost among other high-order ROS methods that we tested.  
\section{Model Problem}\label{s4}

The study of the swimming of an undulating flagellum in a viscous fluid is a classical problem in biological fluid dynamics, beginning with G.I. Taylor’s analysis of the swimming of an infinite sheet or cylinder undergoing small amplitude oscillations \citep{Taylor1951, Taylor1952}. It was shown that for a fixed wavelength, the swimming speed of  the flagellum was second order in amplitude and first order in wave speed. In this simplest class of models, the idealized flagellum followed prescribed kinematics with a traveling wave of a fixed amplitude and wavelength that could not deviate from the prescribed values. However, models that capture the fluid-structure interaction of a flexible flagellum with the surrounding fluid allow the emergent shape of the flagellum to be affected by the viscous load of the fluid. These more detailed models include {\it preferred curvature} models where the flagellum is in pursuit of a preferred shape and generates forces derived from tensile and bending energies. How closely the flagellum adheres to the preferred shape depends upon material parameters such as the tensile stiffness and the bending rigidity. The energy formulation of an undulating, planar flagellum \citep{FAUCI198885} is now a standard {\it preferred curvature } model, and has been used to study flagellar motility in both Newtonian (e.g. \citep{OLSON2011203}) and viscoelastic fluids (e.g. \citep{Teran:2010, Thomases_Guy_2017}). As such, we choose the coupled system of a single, flexible flagellum in an unbounded, three-dimensional Stokes fluid simulated using the standard method of regularized Stokeslets \citep{Cortez05} as the model problem to probe the computational efficiency of the time-integration methods discussed in Section \ref{s3}.

\subsection{Formulation of the ODE System}\label{ss41}
\subsubsection{Centerline representation of filament}
We consider a slender filament $\bm{x}(s,t)$ immersed in a viscous fluid  
that is in pursuit of a planar, traveling sinusoidal wave with 
preferred curvature $\kappa(s,t)$, where $s\in[0, L]$  is the arc length parameter. 
While we will be tracking the position of the slender flagellum using just its centerline, the regularization parameter  $\epsilon$ is taken to be a proxy for flagellar radius (e.g. \cite{Morshed25, Smith09}). 
The filament is initialized along this ideal sine wave, in the plane $z=0$: 
\begin{align}
	\bm{x}(s,0)&= \langle x(s),A\sin(ks),0\rangle,\label{init}
    \end{align}

    \noindent
   and the traveling wave of preferred curvature is: 
    \begin{align}
	\kappa(s,t)&=  \dfrac{-Ak^2\sin(ks+\omega t)}{\sqrt{1-A^2k^2\cos^2(ks+\omega t)}}\label{curv},
\end{align}
where $A, k,\omega$ denote the preferred amplitude, wavenumber, and frequency of the wave, respectively.
The non-dimensional parameters used here are shown in Table \ref{tab01}. 
 
\subsubsection{Forces driving the filament undulation}
At each time step, the position of the filament will be updated according to the ODE system in Eqn. \ref {systemode}. The velocities of the filament points are computed by applying the regularized Stokeslet kernel to the tensile and bending forces defined at each filament point.  We note that the Stokeslet kernel is a function of the current locations of the filament points, as are the forces, which are formulated below.

Given the preferred curvature in Eqn. \ref{curv}, we define the  tensile and bending energies from which the forces are  derived \citep{FAUCI198885, OLSON2011203, CORTEZ2018}: 
\begin{subequations}\label{eq_con}
\begin{align}
	E_{T}&= \dfrac{S_T}{2}\int_{0}^L \bigg(||\dfrac{d\bm{x}(s)}{d s}||-1\bigg)^2 ds,\\
	E_C&= \dfrac{S_C}{2}\int_{0}^L \bigg[\bigg(x'(s)y''(s) - x''(s)y'(s)\bigg)-\kappa(s,t)\bigg]^2 ds.
\end{align} 
\end{subequations}
Note that these energies are always non-negative, and are minimized when the filament has the preferred shape.  How closely this shape is achieved depends upon the tensile stiffness parameter  $S_T$ and the bending rigidity, or curvature stiffness parameter,  $S_C$. 
In discrete form (approximating derivatives by corresponding finite difference operators and utilizing a quadrature rule for integral approximation), we obtain the following, evaluating forces at discrete nodes $\bm{x}_k=(x_k,y_k,z_k)$ \citep{CORTEZ2018}:
\begin{subequations}\label{eq_ds}
\begin{align}
	\Tilde{E_{T}}&= \dfrac{S_T}{2}\sum_{k=1}^{N-1} \bigg(\dfrac{||\bm{x_{k+1}-x_{k}||}}{\Delta s}-1\bigg)^2 \Delta s\label{e41},\\
	\Tilde{E_{C}}&= \dfrac{S_C}{2}\sum_{k=2}^{N-1} \bigg[(D^0 x_k D^2 y_k - D^2 x_k D^0 y_k)-\kappa_k\bigg]^2 \Delta s\label{e42},\\
    \bm{\Tilde{f}}_{k}&= -\dfrac{\partial}{\partial \bm{x}_{k}}( \Tilde{E_{T}}+ \Tilde{E_{C}}),\label{efd}
\end{align}
\end{subequations}
where $D^0$ and $D^2$ are central finite difference operators for the first and second derivatives. Moreover, $N$ represents the number of discrete nodes along the filament and $\Delta s$ is the spacing between nodes. We note that because we are considering a single flagellum in free-space, initialized in the $z=0$ plane, the forces are confined to that plane, and thus the flagellum remains in that plane as time evolves. 

With the above formulation of forces, we can update the filament's position by solving the 
ODE system in Eqn. \ref {systemode}.
The velocity $\bm{u}(\bm{x}_j)$ at the node $j$ on the filament requires summing the contributions from all $N$  nodes via the regularized Stokeslet matrix.
We note that in this work we use the direct $\mathcal{O}(N^2)$ summation to evaluate the velocity at filament nodes, but other implementations of MRS have used fast methods, such as kernel-independent fast multipole methods (e.g. \cite{ROSTAMI201660,LAGRONE201973}).

Table \ref{tab01} shows the non-dimensional parameters used here that are based upon those derived for a mammalian flagellum in \citep{OLSON2011203, SIMONS201481}. For the choice of $S_C = 0.05$ and $\epsilon = 0.01$ in the table (the value of $\epsilon$ is discussed in Section~\ref{modelparameters}), the corresponding dimensional variables represent
a flagellum of length $\Tilde L = 100 \mu m$,  beat frequency of $10$ Hz, flagellar radius  $\Tilde \epsilon = 1 \mu m$, and bending rigidity of $EI = 10^{-21} N m^2$ undulating in water.  In these elastohydrodynamic systems of flexible fibers in flow, the key nondimensional parameter that measures the relative importance of viscous forces to elastic forces is the ``Sperm'' number (e.g. \cite {GGSK2010, gallagher2023, Morshed25}): 

  $$Sp = \left(\displaystyle\frac{4 \pi \Tilde\mu \Tilde \omega \Tilde L^4}{ (\ln (2 \Tilde L /\Tilde \epsilon + \frac {1}{2} )EI\
} \right)^{1/4}.$$ 

\noindent
Here we see that when considering just the bending rigidity, stiffer filaments (larger $EI$) lead to smaller $Sp$, whereas softer filaments (smaller $EI$) give rise to larger $Sp$.  We note that the bending stiffness $S_C$ scales linearly with $EI$ (\cite{OLSON2011203}).  For our choice of $S_C$=.05 the resulting sperm number is $Sp \approx 11$, while for the stiffer case $S_C$ = .2 the resulting sperm number is $Sp \approx 8$.  The typical range exhibited by mammalian sperm flagella range between 4 and 24 (\cite {GGSK2010}).  Note that while $S_C$ =.2 corresponds to a bending rigidity of $EI = 4 \times 10^{-21} Nm^2$ (a value attributed to bull sperm (\cite{bullsperm})), the bending rigidity measured for rat sperm is two orders of magnitude larger ($EI = 10^{-19} Nm^2$) ( \cite{ratsperm}). 

\begin{table}[H]
  	\centering
  	\begin{tabular}{|c|c|c|}
  		\hline
  		Description& Notation &Value \\ \hline
        Fluid Viscosity   & $\mu$ & 1\\
         Filament Length  & $L$ & 1\\
  		Amplitude & A & 0.1 \\ 
        Beat frequency &$\omega$ & $2\pi$\\
  		Wave number &$k$ & $2\pi$\\
        Curvature stiffness constant&$S_C$&$0.05$ or $0.2$\\
  		Tensile stiffness constant &$S_T$& 100\\ 
  		Flagellar radius &$\epsilon$ & .0071 or .01 \\ \hline
  	\end{tabular}
  	\caption{Dimensionless parameter values considered in this article.}
  	\label{tab01}
  \end{table}
  
\begin{figure}[H]
	\centering
	\includegraphics[width=0.7\textwidth,height=4 in]{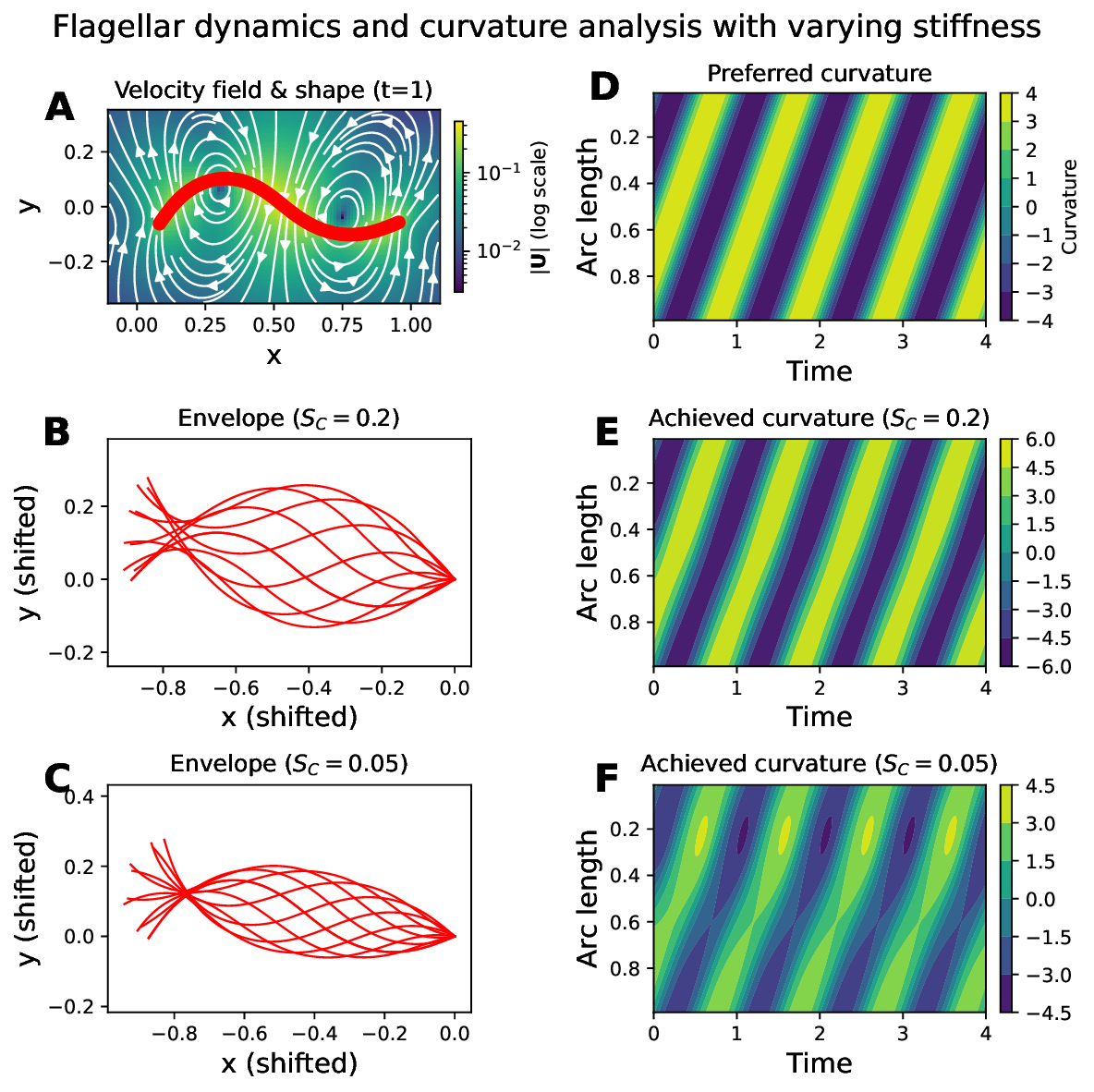} 
	\caption{(\textbf{A}) Streamline of the flow together with velocity magnitude and swimming filament at $t=1$, (\textbf{B}) envelope with $(S_C,\epsilon)=(0.2,0.01)$ 
    (\textbf{C}) envelope with $(S_C,\epsilon)=(0.05,0.01)$ 
    (\textbf{D}) preferred curvature, (\textbf{E}) achieved curvature with $(S_C,\epsilon)=(0.2,0.01)$, and (\textbf{F}) achieved curvature with $(S_C,\epsilon)=(0.05,0.01)$. For time integration, we have used \text{RO$\widetilde{W}$3} with $\Delta t=0.5\times10^{-2}$.  The number of discrete points on the flagellum was $N=101$.}  
	\label{fig01}
\end{figure}
\subsection{Visualization of Velocity Field and Curvature}
Figure \ref{fig01}A shows a snapshot of the model flagellum after it has completed one beat period ($t=1$) and the surrounding streamlines in the plane of beating. 
Figure \ref{fig01}B shows a kymograph of the preferred wave of 
curvature specified to move along the flagellum over four temporal periods. Labeling the leftmost point on the flagellum as $s=0$ and the rightmost point as $s=1$, this preferred curvature wave travels to the left along the body, and the overall swimming direction will be to the right.  Figures \ref{fig01}B, C show the envelopes of the emergent waveforms in two distinct simulations:  $S_C$ = 0.2 and $S_C$ = 0.05, respectively, and Figures \ref{fig01}E, F show the emergent curvature kymographs.  In Figures \ref{fig01}B, C, images of the flagellum are shown at ten equispaced times within the third period of beating, where the heads' positions were translated to the origin. 
We see that for the larger bending stiffness, the achieved curvatures track more closely the preferred curvatures, and the flagellum achieves a larger amplitude beatform. 
In fact, for $S_C=0.05$ the relative difference between achieved and preferred curvatures (computed using the norm $\frac{||\kappa_{\text{preferred}}-\kappa_{\text{achieved}}||_{2}}{||\kappa_{\text{preferred}}||_{2}}$, where $||\cdot||_2$ denotes the Frobenius norm) over time $t\in[0,4]$ was 0.24, while for $S_C=0.2$ this difference decreased to 0.08.

\subsection{Model Parameters}\label{modelparameters}

The elastohydrodynamic problem of a flexible filament in  Stokes flow involves several physical parameters -  those chosen for our model swimmer are listed in Table \ref{tab01}.  In this work, we focus on two parameters that have added to the computational challenges when using realistic values in simulations: bending stiffness and filament thickness. 

\textbf{Bending stiffness.}
The bending stiffness constant determines how closely the filament achieves its preferred curvature, and the tensile stiffness constant determines how closely inextensibility is maintained  \citep{SIMONS201481}.  As discussed above, reference values for bending rigidity are available for different sperm flagella, and span several orders of magnitude.
Increasing the filament stiffness parameters require smaller time steps for ERK methods to remain stable \citep{Nourian_Shum_2023, makanga23}. Of these two stiffness constants, bending stiffness has a more significant influence on the critical time step than tensile stiffness \citep{JABBARZADEH2020, makanga23, ferranti2025}. In particular, the bending stiffness is an important physiological parameter that is central to modeling studies.  As such, below we will investigate how it influences the numerical stiffness of the governing ODE system. 

\textbf{Filaments' Thickness or Radius.} In MRS, the fluid motion, computed with Eqn.~\eqref{systemode}, depends on $\epsilon$ and on the node forces, which depend on the node separation, $\Delta s$. 
In practice, choosing  $\Delta s$ to be 
approximately equal to $\epsilon$ or smaller is best for accuracy and to avoid fluid leakage between nodes~\citep{CORTEZ2018, Carichinoetal2021}. Thus in this work, we choose the number of nodes $N$ along the flagellum  such that $\Delta s = \epsilon$. In applications of cilia and flagella that track the centerline of the filament, the regularization parameter $\epsilon$ may be interpreted as the filament radius~\citep{CORTEZ2018, Carichinoetal2021, Smith09, Walker_et_al_2019}. In this case, within this MRS model, the number $N$ of nodes is larger for model flagella with smaller radius.   An alternative method, which is not pursued here, is to use the method of regularized Stokeslet segments~\citep{CORTEZ2018, Walker_et_al_2019} instead of MRS. In this method, the flagellum is represented as a piecewise linear curve with polynomial forces on each segment. The segment length is decoupled from the regularization parameter $\epsilon$ and chosen based on curvature and force variation along the flagellum. Nevertheless, the smallness of $\epsilon$ leads to time-step restrictions.  For perspective, we note that a human sperm flagellum has a length of approximately 60 $\mu m$ and a flagellar radius of .085 - .44 $\mu m$ \citep{humansperm}, while a mouse sperm flagellum has a length of approximately 100$ \mu m$ and a flagellar radius of .5 - 1 $\mu m$ \citep{micesperm-radius}.  Therefore, the aspect ratio of flagellar radius to flagellar length $\epsilon / L$ used here falls within the physiological range (see Table ~\ref{tab01}). 

\section{Effect of Parameters on ERK's Efficiency}\label{s5}
Explicit Runge-Kutta (ERK) methods require careful selection of the step size to ensure stability. In this section, we characterize the step size restrictions of ERK methods as a function of model parameters discussed in Section \ref{s4}.
Using linear stability analysis, we will show that the step size of an ERK method is restricted by
\begin{align}
    \Delta t_{max} =\mathcal{O}\left(\frac{\epsilon^4}{S_C}\right).
\end{align}
This implies that highly stiff and slender filaments are the most numerically challenging to integrate with an ERK method.
\subsection{Linear Stability}\label{ss51}
The Dahlquist test equation $x'=\lambda x$ is commonly used to study a method's stability properties \citep{wanner1996}. When an ERK method is used to solve the Dahlquist equation, the computation reduces to a recurrence $x_{n+1}=R(z)x_{n}$, where $R(z)$ is known as the stability function. The stability region of a method contains all the points $z = \lambda \Delta t$ such that the method produces a bounded solution. It is defined as: 
\begin{align}
    S =\{z\in\mathbb{C}: |R(z)| \le 1\}.
\end{align}
For example, Euler's method has the stability function $R(z) = 1 + z$, hence its step size is restricted by: $ \Delta t\le\frac{2}{ |\lambda|}$.

This concept can be extended to study the stability of linear systems $\bm{x}' = B\bm{x}$, where $B$ is a square, diagonalizable matrix. An ERK method produces a bounded solution if all the scaled eigenvalues $z_i = \lambda_i (B) \Delta t$ lie within the method's stability region. For a nonlinear system, we can approximate stability restrictions by considering the eigenvalues of the Jacobian $J = \frac{\partial \bm{u}}{\partial \bm{x}}$ \citep{wanner1996}. The step size $\Delta t$ of a time integrator for such a system is then restricted by the requirement that all $\lambda_i(J) \Delta t $ remain strictly within the stability region of the method throughout the integration interval.

\subsection{Spectrum of the Discretized Model Problem \eqref{eq_ds}}
To analyze the stability restrictions of an ERK method, we first need to understand the spectrum of our discrete system \eqref{eq_ds} as a function of the model parameters. Pairing this information with the stability region of an ERK method, then allows us to determine the step size restrictions for that method.
\subsubsection{The Discrete Spectrum}
\begin{figure}[H]
    \centering
    \begin{subfigure}{0.45\textwidth}
        \centering
        \includegraphics[height=4 cm ]{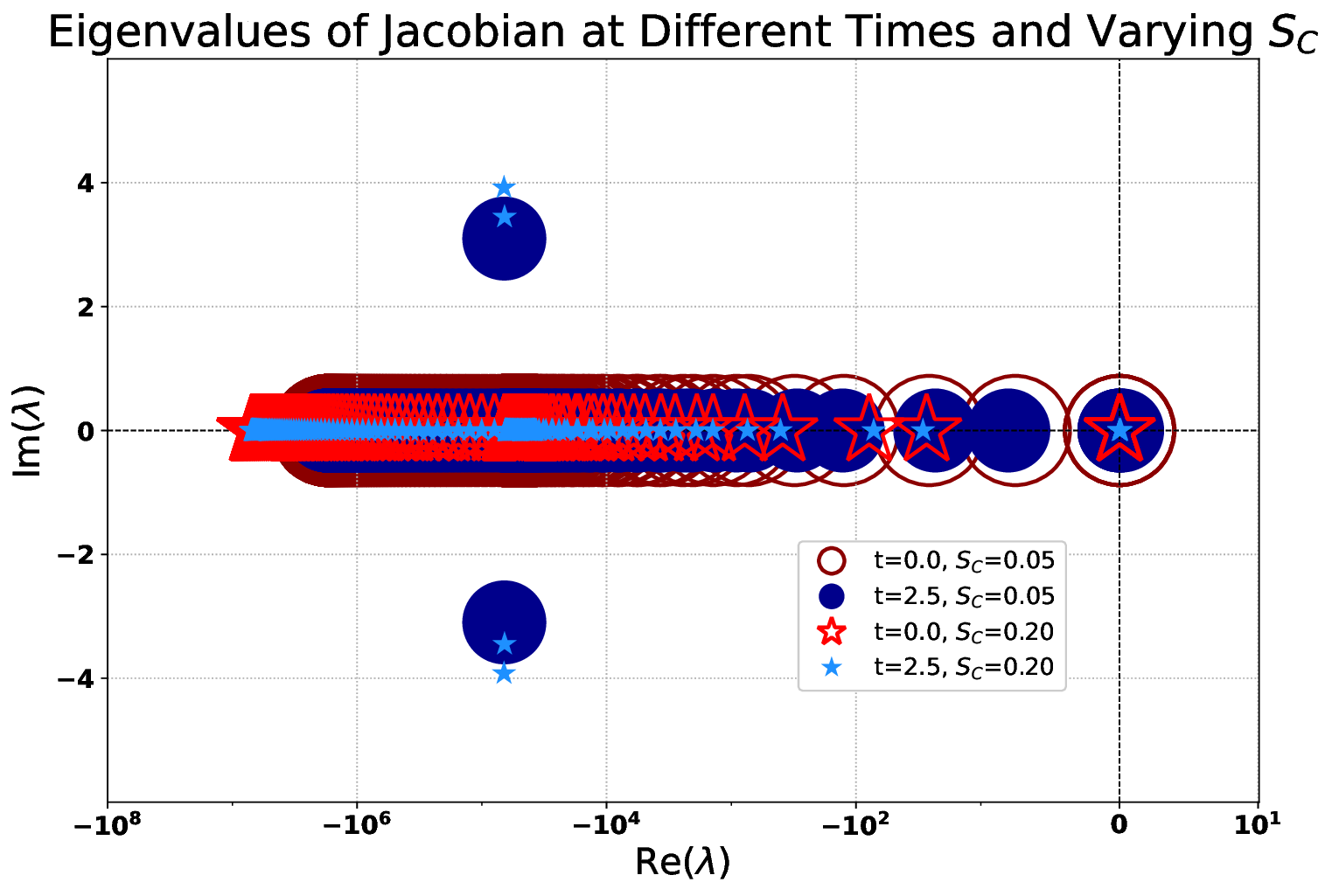}
        \caption{Jacobian eigenvalues with varying $S_C$}
        \label{fig2aU}
    \end{subfigure}
    \begin{subfigure}{0.45\textwidth}
        \centering
        \includegraphics[height=4 cm ]{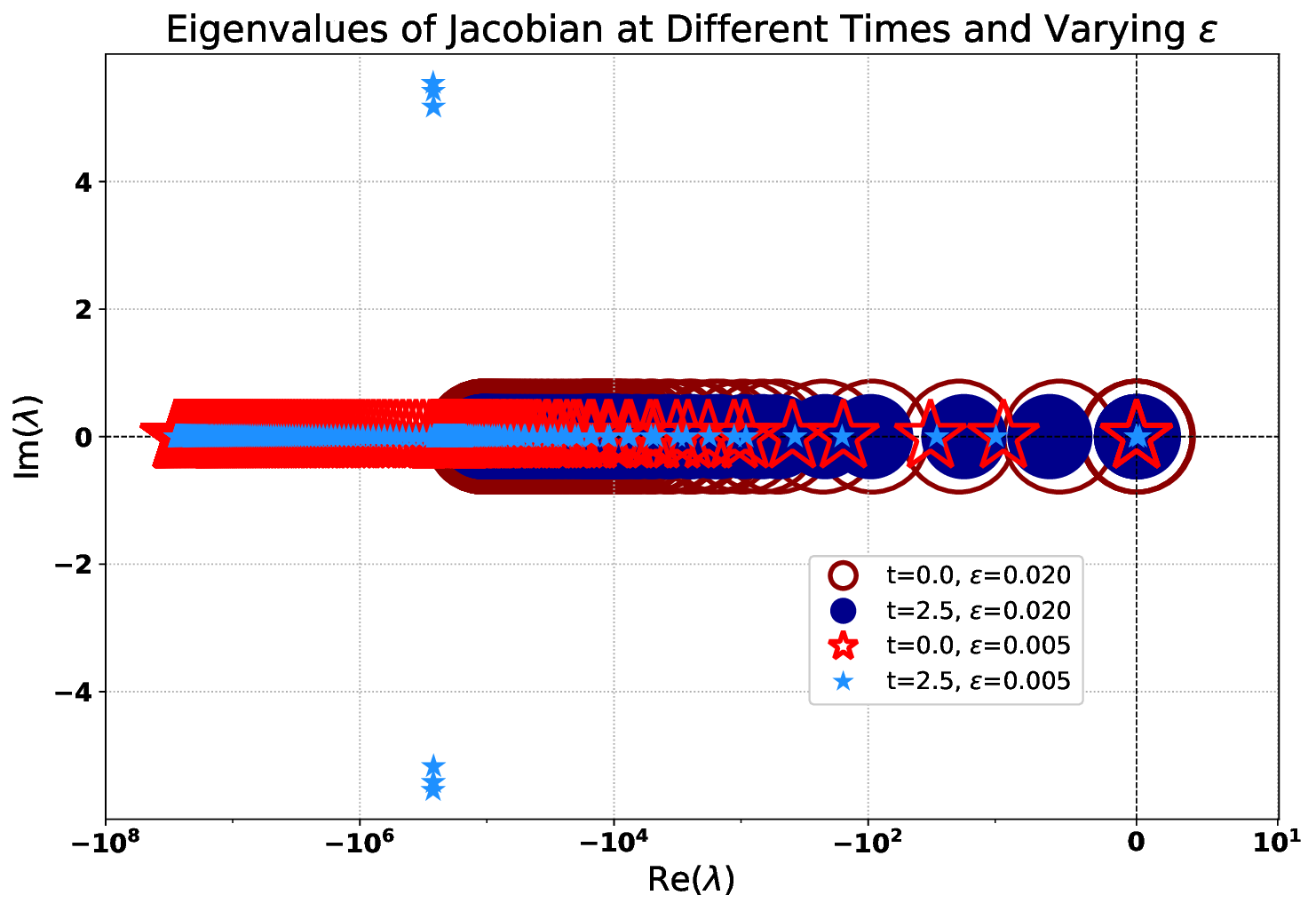}
        \caption{Jacobian eigenvalues with varying $\epsilon$}
        \label{fig2bU}
    \end{subfigure}
    \caption{Eigenvalues of the right-hand Jacobian considering different parameters at the initial configuration and after  2.5 temporal beat periods: (a) with $S_C\in\{0.05, 0.2\}$ and $ \epsilon=0.01$, (b) with $S_C =0.05$ and $\epsilon\in\{ 0.02, 0.005\}$. In both figures that the $x$-axis span $[-10^{-8},~ 10]$ while the $y$-axis spans  $[-6,6]$; hence, for all the parameters, the eigenvalues lie very close to the negative real axis.}
    \label{fig2U}
\end{figure}

In Figure \ref{fig2U}, we plot the eigenvalues of the Jacobian $\frac{\partial \bm{u}}{\partial\bm{x}}(\bm{x}(t))$ at $t \in\{ 0, 2.5\}$ for various stiffness constants and thicknesses. Specifically, in Figure \ref{fig2aU}, we plot  the spectrum for bending stiffness constants $S_C\in\{0.05, 0.2\}$ and thickness $\epsilon=0.01$. In Figure \ref{fig2bU}, we plot spectrum for filament thicknesses $\epsilon\in\{ 0.02, 0.005\}$ and bending stiffness $S_C=0.05$. To compute eigenvalues, we used \text{RO$\widetilde{W}3$} to obtain the solution $\bm{x}(t)$, and used automatic differentiation in JAX \citep{jax2018github} for computing the Jacobians.

The eigenvalue plots indicate that all eigenvalues lie almost entirely along the negative real axis. Moreover,  the real parts remain identical over time. This indicates that the system behaves like a diffusive system rather than an advective one.  
 \subsubsection{Spectral Radius as a Function of Model Parameters}

Here, we characterize how the spectral radius of the Jacobian, defined as $\rho(J(\bm{x}(t)))=\max_i|\lambda_i(t) |$, is affected by the model parameters $S_C$ and $\epsilon$. Interestingly, we observe that the spectral radius is almost completely independent of time $t$. This phenomenon is also visible in Figure \ref{fig2U}, where the dominant real eigenvalue remains nearly identical for different $t$ values. Therefore, in the plots below, we show the spectral radii solely at the initial time $t=0$.

\begin{figure}[H]
    \centering
    \includegraphics[width=0.85\linewidth,height=5 cm ]{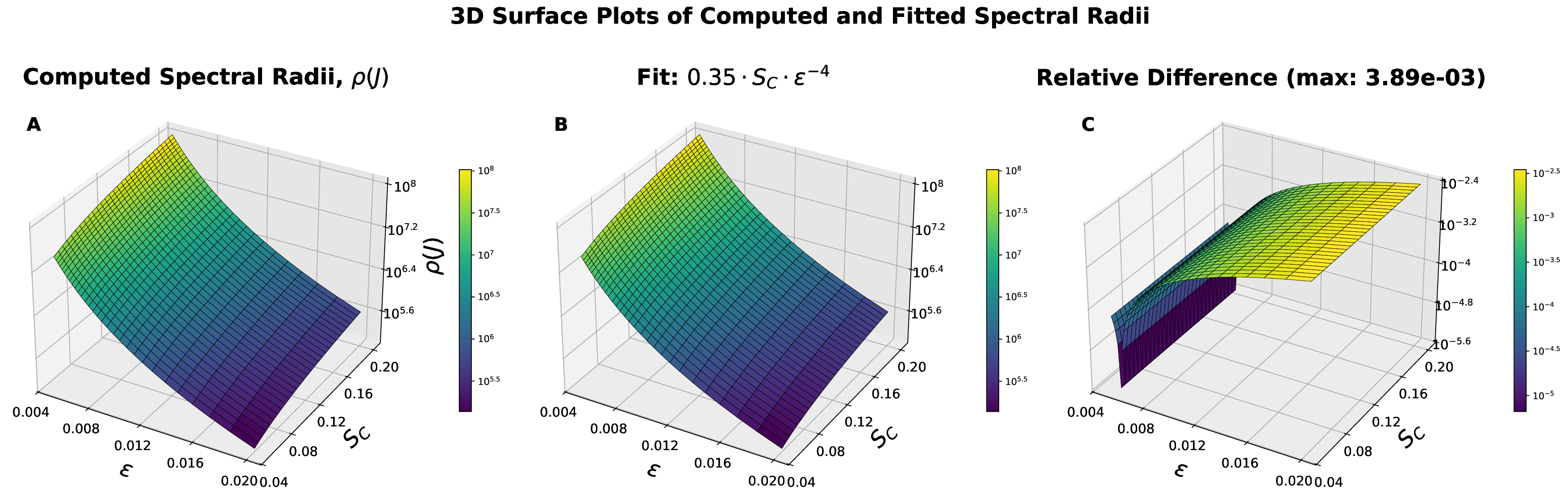}
    \caption{(\textbf{A}) Surface plot for spectral radii ($\rho(J)$) of the Jacobian considering varying bending stiffness $S_C$ and thickness $\epsilon$ computed at the initial configuration, (\textbf{B}) fitted spectral radii  ($\rho(J)$) of the Jacobian, and (\textbf{C}) relative difference between computed and fitted spectral radii of the Jacobian.}
    \label{fig13s}
\end{figure}

     In Figure \ref{fig13s}\textbf{A}, we plotted $\rho(J(\bm{x}(0)))$ considering $50$ bending stiffness constants $S_C\in\{0.05,0.2\}$ and filament thicknesses $\epsilon \in\{0.005,0.02\}$. Figure \ref{fig13s}\textbf{B} represents a fitted plot for spectral radii. We observe linear scaling of $S_C$ as $S_C$ increases, and inverse quartic scaling as we decrease $\epsilon$. In particular, we can approximate $\rho (J)= 0.35 \cdot S_C \cdot \epsilon^{-4}$. This gives a good estimate with a maximum relative difference of about $3.88 \times 10^{-3}$ as visualized in figure \ref{fig13s}\textbf{C}.

Based on our results, we estimate the dominant eigenvalue approximately as 
\begin{align}\label{e18}
    \rho\left(J(\bm{x}(t))\right)\approx  \frac{7}{20} \frac{S_C}{\epsilon^4}=\mathcal{O}\left(\frac{S_C}{\epsilon^4}\right).
\end{align}
We note that the inverse quartic scaling in $\epsilon$ has been observed in previous studies for similar biofluid models \citep{JABBARZADEH2020, makanga23, ferranti2025}.

\subsection{Step Size Restrictions on ERK Methods Depending on Parameters}
By combining the spectral radius of the Jacobian \eqref{e18} with the fact that the discrete spectrum is almost purely diffusive, we can estimate the largest stable step size of an ERK method via
\begin{align}\label{e19}
    \Delta t_{\text{max}} = \frac{D}{\rho (J  )}=\mathcal{O}\left(\frac{\epsilon^4}{S_C}\right)
    \quad \text{where} \quad
    \renewcommand\arraystretch{1.25}
    \begin{tabular}{c|cc}
          & RK1 & RK3 \\ \hline
        D& 2   & 2.52 
    \end{tabular}.
\end{align}
This bound ensures that all the eigenvalues of the Jacobian are inside the method's stability regions. We validate these bounds using two numerical experiments run to final time $t=1$. In the first, we consider two different bending stiffnesses $S_C\in\{0.05,0.2\}$ and a thickness of $\epsilon=0.01$; results are shown in Table \ref{tab03c}. In the second case, we consider two different thicknesses $\epsilon\in\{0.01,0.0071\}$ and a bending stiffness of $S_C=0.05$; results are shown in Table \ref{tab03}.
 Our observations indicate that the estimated step sizes are excellent approximations of the real values.  We note that the actual $\Delta t_{\text{max}}$  were calculated using a binary search.

\begin{table}[H]
	
	\centering
	\begin{tabular}{|c|c|c|c|c|}
		\hline
		Method  & Estimated  $\Delta t_{\text{max}} $ & Actual $\Delta t_{\text{max}} $    &Estimated $\Delta t_{\text{max}} $    &Actual $\Delta t_{\text{max}} $  \\	  &  $ (S_C=0.05)$ &  $ (S_C=0.05)$  &  $ (S_C=0.2)$   &  $ (S_C=0.2)$  \\
		\hline
		RK1   & $1.152e-06$ &   $1.175e-06$& $2.881e-07$ & $2.919e-07$ \\
		\hline
		RK3  &$1.452e-06$  &  $1.463e-06$ & $3.63e-07$ & $3.657e-07$ \\
		\hline
	\end{tabular}  
	\caption{Step size restrictions for RK1 and RK3  based on spectrum analyses considering $\epsilon=0.01$ and two different $S_C$ values. Actual  $\Delta t_{\text{max}} $ are based on simulation and  estimated  $\Delta t_{\text{max}} $ are based on equation \eqref{e19}.}
	\label{tab03c}
\end{table}

 \begin{table}[H]
	
	\centering
	\begin{tabular}{|c|c|c|c|c|}
		\hline
		Method  & Estimated  $\Delta t_{\text{max}} $  & Actual $\Delta t_{\text{max}} $    &Estimated $\Delta t_{\text{max}} $    &Actual $\Delta t_{\text{max}} $  \\	  &  $ (\epsilon=0.01)$ &  $ (\epsilon=0.01)$  &  $  (\epsilon=0.0071)$   &  $  (\epsilon=0.0071)$  \\
		\hline
		RK1   & $1.152e-06$ &   $1.175e-06$& $2.999e-07$ & $3.03e-07$ \\
		\hline
		RK3  &$1.452e-06$  &  $1.463e-06$ & $3.778e-07$ & $3.792e-07$ \\
		\hline
	\end{tabular}  
	\caption{Step size restrictions for RK1 and RK3 based on spectrum analyses considering $S_C=0.05$ and two different $\epsilon$ values. Actual  $\Delta t_{\text{max}} $ are based on simulation and  estimated  $\Delta t_{\text{max}} $ are based on equation \eqref{e19}.}
	\label{tab03}
\end{table}

\section{Asymptotic Speedup of a ROS Method}\label{s6}

Using the step size restrictions for ERK methods derived in Section \ref{s5}, we can calculate asymptotic runtime for ERK and ROS methods as a function of model parameters. We estimate the computational cost of a time integrator by multiplying the per-step work by the number of steps required to obtain a stable solution.  

\subsection{Computational Cost of an ERK Method}\label{ss61}

The cost of evaluating the right-hand side of the ODE system \ref{e11} is $\mathcal{O}(\epsilon^{-2})$ (see Section \ref{s4}), and an $s$-stage ERK method requires $s$ function evaluations per time step. Using the estimate of the spectral radius $\rho(J )$ given by equation \eqref{e18}, the number of steps required to compute a stable solution at final time $t_f$ by an ERK method is $\mathcal{O}( t_f \cdot S_C \cdot \epsilon^{-4} )$. As a result, the computational cost of an $s$-stage ERK method to run up to  final time $t_f$ is 
 \begin{align}\label{e20}
   \text{Cost}_{\text{ERK}}=  \mathcal{O}( s\cdot t_f\cdot S_C\cdot\epsilon^{-6} ).
 \end{align}
\subsection{Computational Cost of a ROS Method}\label{ss62}

 The computational cost for a ROS method is determined by function evaluations, Jacobian construction, and linear solves. The cost of function evaluations is $\mathcal{O}(\epsilon^{-2})$, and an $s$-stage ROS method requires $s$ function evaluations.
Constructing a Jacobian incurs a cost of $\mathcal{O}(\epsilon^{-3})$ (recall that the dimension of the system $N=\mathcal{O}(\epsilon^{-1})$ and $N$ function evaluations are required to build the Jacobian). If a direct solver is used for solving linear systems, then the cost of matrix decomposition (for example, decomposition of $\bm{I}- \Delta t\bm{J}_n$ in equation \ref{ROS1} ) is $\mathcal{O}(\epsilon^{-3})$.  The cost of back substitution for an $s$-stage ROS method is $\mathcal{O}(s\cdot\epsilon^{-2})$.
Lastly, all ROS, ROW,  and RO$\widetilde{\text{W}}$ methods we consider in this article are $L$-stable and thus have no theoretical restrictions on their step sizes due to linear stability. We therefore assume that the number of steps required for a stable solution is $\mathcal{O}(t_f)$.
 As a result, the computational cost of an $s$-stage ROS  method to run up to  final time $t_f$ is 
 \begin{align}\label{e21c}
   \text{Cost}_{\text{ROS}}=  \mathcal{O}(t_f)\cdot\left(\underbrace{\mathcal{O}( s\epsilon^{-2} )}_{\bm{u} \text{ Eval}} +\underbrace{\mathcal{O}( \epsilon^{-3} )}_{\text{Jacobian}}+\underbrace{\mathcal{O}( \epsilon^{-3} )}_{\text{Decomposition}}+\underbrace{\mathcal{O}( s\epsilon^{-2} )}_{\text{Solve}}\right)\approx      \mathcal{O}(  t_f\cdot \epsilon^{-3} ).
 \end{align}


\subsection{Quantifying Computational Costs}

\begin{figure}[H]
    \centering
    \includegraphics[width=0.7\linewidth,height=2.5in]{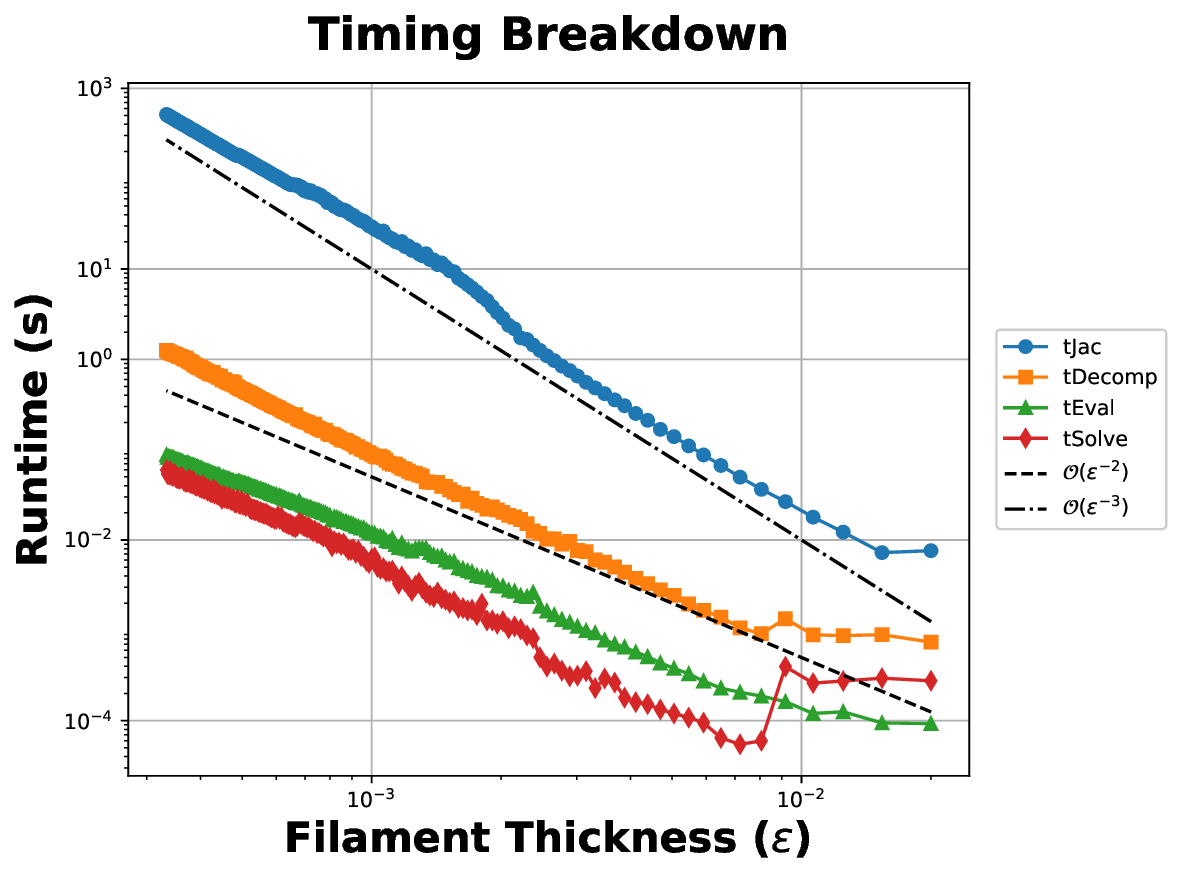}
    \caption{Timing breakdown for various costs associated with ERK and ROS methods, considering varying thickness $\epsilon$ and $S_C=0.05$.}
    \label{fig_tb}
\end{figure}

In Figure \ref{fig_tb}, we quantify the real-world computational costs of the operations associated with the ERK and Rosenbrock methods. These include Jacobian construction, function evaluations, LU matrix decomposition, and backward substitution to solve linear systems. For each operation, we also approximate asymptotic runtimes as a function of $\epsilon$. Note that we do not consider $S_C$ because the computational cost of all operations is independent of this quantity.

All asymptotic runtimes agree closely with our estimates, with the exception of LU decomposition, which scales as $\mathcal{O}(\epsilon^{-2 })$ instead of $\mathcal{O}(\epsilon^{-3})$ for small matrices. The expected scaling only appears once the matrix dimension exceeds approximately 5000, equivalent to 2500 nodes on the filament. Another important observation is that the Jacobian construction cost dominates the other costs; in \cref{SS72} we will discuss strategies for improving this computational bottleneck.

\section{Numerical Experiments and Discussions}\label{s7}
In this section, we test the efficiency of the time integrators in Table \ref{tabm} for integrating the flexible swimming filament from Section \ref{s4}.
To assess how efficiency changes across different values of $S_c$ and $\epsilon$, we consider the three parameter sets described in \cref{tab:numericalexperimentparameters}. The remaining parameters are listed in \ref{tab01}, and all simulations are run up to time $t=1$. 

\begin{table}[h!]
    \centering
    \begin{tabular}{r|cc}
        & $S_c$ & $\epsilon$ \\ \hline
        base case & 0.05 & 0.01 \\
        stiff filament & 0.2 & 0.01 \\
        slender filament & 0.05 & 0.0071
    \end{tabular}
    \caption{$(S_c, \epsilon)$ parameter pairs for numerical experiments. Compared to the base case, the stiff filament has increased $S_c$, while the slender filament has smaller $\epsilon$.}
    \label{tab:numericalexperimentparameters}
\end{table}

As there is no exact solution for the system under consideration, we computed reference solutions $\bm{x}_{ref}$ using the RK45 \citep{DORMAND198019} integrator with a small error tolerance. We compute the error at the final time using the norm $\frac{\lvert\lvert\bm{x}_{ref}-\bm{x}_{num}\rvert\rvert_{\infty}}{ 2A} $, where $A$ is the amplitude of the filament. We normalize the maximum spatial error by the preferred peak-to-peak amplitude ($2A$) to measure relative numerical deviations against the filament's transverse beating envelope.  Lastly, all the codes used to run the experiments in this section are available in \cite{GIT_REP}. 

\subsection{ROS Methods Using Finite Difference Jacobians}\label{ss71}
\begin{figure}[H]
    \begin{subfigure}{\textwidth}
         \includegraphics[width=\textwidth,height=2.4in]{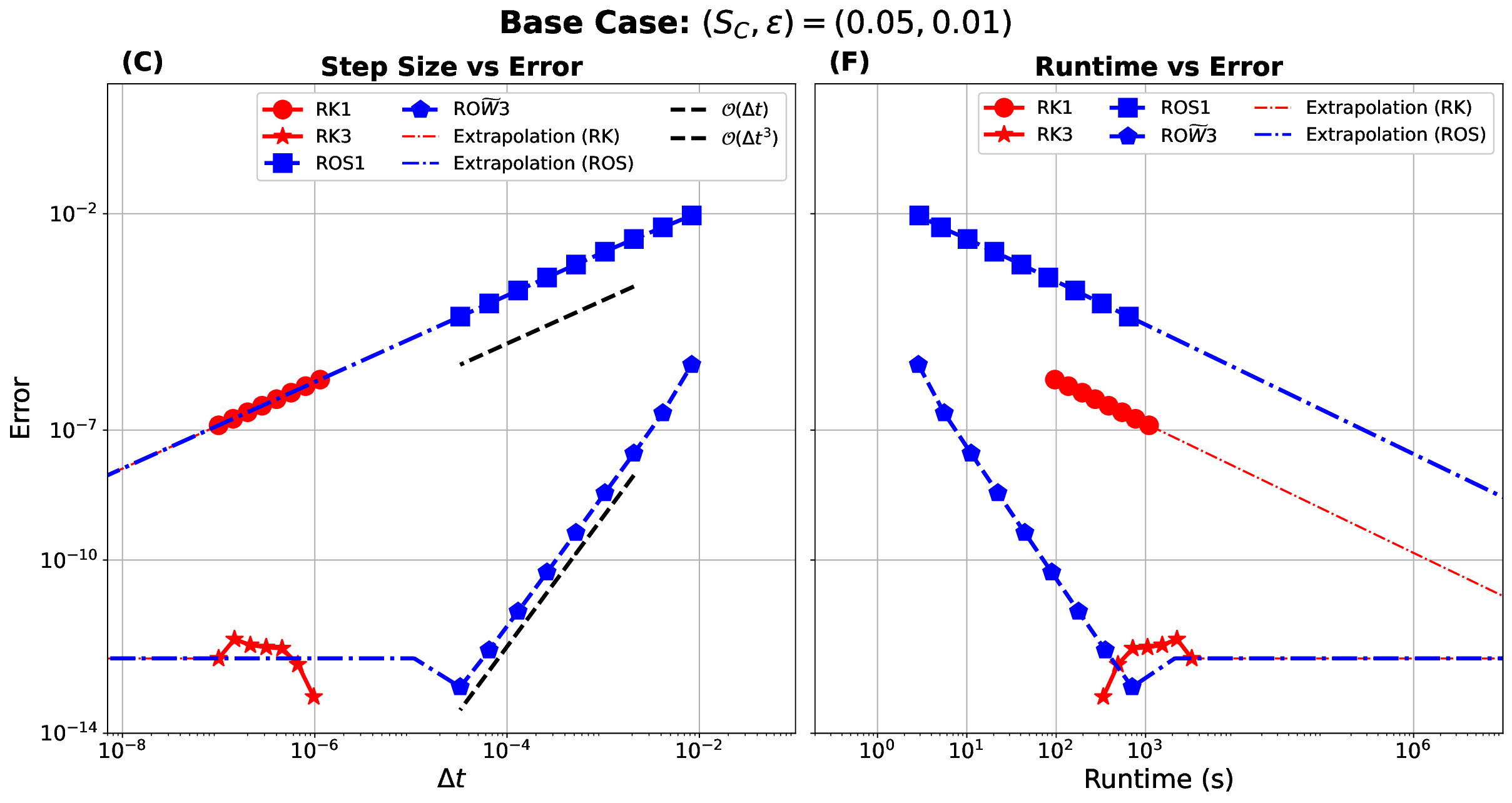}
     \end{subfigure}
     \hfill
     \begin{subfigure}{\textwidth}
         \includegraphics[width=\textwidth,height=2.4in]{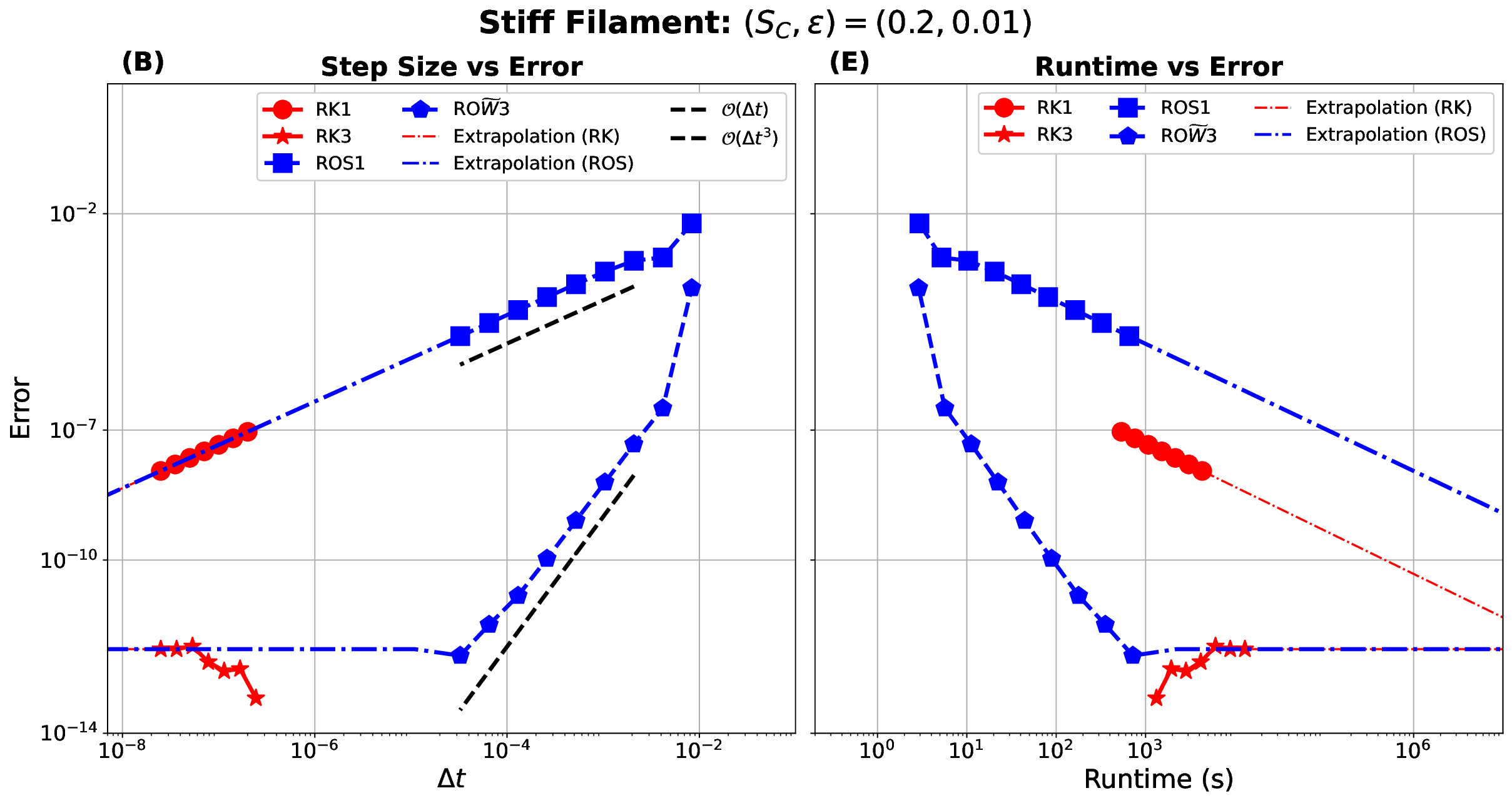}
     \end{subfigure}
     \hfill
         \begin{subfigure}{\textwidth}
         \includegraphics[width=\textwidth,height=2.4in]{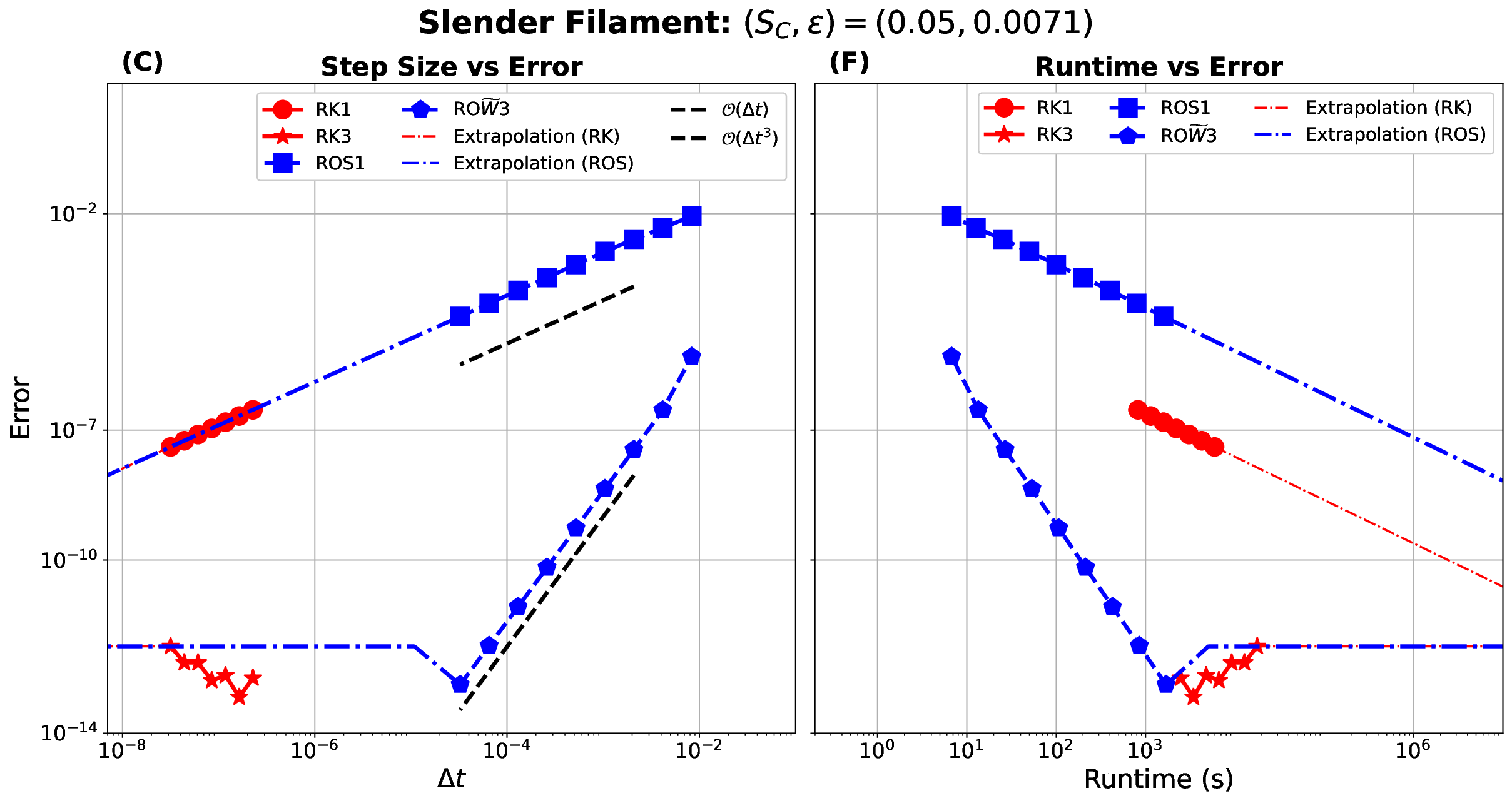}
     \end{subfigure}
     
     \caption{Convergence diagrams (left column) and work-precision diagrams (right column) for the $S_C,\epsilon$ from. The extrapolated dashed-dotted lines predict behavior for step sizes that were too expensive to simulate.} 
      \label{fig17}
\end{figure}

We first compare the computational efficiency of RK1 and RK3 to ROS1 and \text{RO$\widetilde{W}3$} with Jacobians that are computed using the FD approximation \eqref{e47}. 
 Figure \ref{fig17} contains convergence diagrams (step size $\Delta t$ versus errors) and work-precision diagrams (runtime versus errors) for ROS and ERK methods. 
 
    \textbf{Convergence.} 
Figures \ref{fig17}(A), \ref{fig17}(B), and \ref{fig17}(C) show the convergence plots for base, stiff, and slender filaments, respectively. Our results indicate that ROS methods are able to produce stable solutions for $\Delta t \approx 10^{-2}$ across all parameter values. In contrast, ERK methods require $\Delta t \approx 10^{-6}$ to obtain a stable solution for the base case \ref{fig17}(A) and substantially smaller step sizes for the numerically stiffer parameters \ref{fig17}(B) and \ref{fig17}(C). Between ROS1 and \text{RO$\widetilde{W}3$} methods, ROS1 has improved stability and allows for a maximum step size that is 4 times larger than that for \text{RO$\widetilde{W}3$}.

Lastly, all methods achieved their expected order of convergence up to the point when  machine precision is reached. For RK3, this effect is especially pronounced since the method has already converged to machine precision at the largest stable stepsize.

\textbf{Efficiency.}
Figures \ref{fig17}\textbf{(D)}, \ref{fig17}\textbf{(E)}, and \ref{fig17}\textbf{(F)} show the work-precision plots for base, stiff, and slender filaments, respectively. Overall, ROS methods are significantly more efficient than ERK methods across all three parameters (base, stiff, and slender) and across all error ranges. ROS1 should be preferred for single-digit accuracy (we can achieve around 1--2-digit accuracy within a few seconds), while \text{RO$\widetilde{W}3$} should be preferred for high precision beyond $10^{-2}$. Consistent with our asymptotic analysis, the runtimes for ROS methods in the stiff and base cases are nearly identical, while the slender case takes approximately 3 times longer (since $\epsilon \to \epsilon/1.4$, we expect an increase in computational time of $1.4^3 \approx 2.74$ from \cref{e21c}). Both the seven-stage $RO\tilde{W}3$ method and the one-stage ROW1 method require similar simulation times for the same timestep, which highlights that Jacobian construction, rather than the linear solve required for each stage, is the dominant cost.

For ERK methods, efficiency degrades as the fiber becomes stiffer or more slender. Specifically, the stiff case requires four times the computational work of the base case, while the slender case requires seven times the computational work. These numbers are consistent with our asymptotic formula \cref{e20}, which predicts linear scaling in $S_c$ and inverse hexagonal scaling in $\epsilon$. Among all the integrators, RK1 is the least preferred since it returns solutions with $10^{-6}$ error while taking 20--100 times longer than RO$\widetilde{W}3$. Similarly, RK3 should only be considered if machine precision is required (which is atypical for a modeling simulation).
 
\subsection{Further Enhancements for Rosenbrock methods}\label{SS72}
This subsection contains three strategies that further improve the performance of ROS methods by reducing the computational cost of Jacobian computation (the dominant cost per step as shown in Figure \ref{fig_tb}). 

\subsubsection{Sparse Updates for FD Jacobian Construction}\label{ss721}

We can reduce the cost of Jacobian construction based on the FD formula \cref{e47} by using the evaluation of $\bm{u}(\bm{x})$ to efficiently calculate $\bm{u}(\bm{x}+h \bm{e}_i)$, for $i = 1, \ldots, 3N$. The term $\bm{u}(\bm{x}+h \bm{e}_i)$ represents the velocity field when one node is perturbed in a single dimension $x$, $y$, or $z$. The key observation is that the $i$th perturbed velocity fields can then be written as the sum of $\bm{u}(\bm{x})$ and terms involving a sparse matrix $\Delta \bm{S}^{(i)}$ and vector $\Delta \bm{f}^{(i)}$ via
\begin{align}
    \text{unperturbed velocity:} \quad & \bm{u}(\bm{x}) = \underbrace{\bm{S} \bm{f}}_{\mathcal{O}(\epsilon^{-2})\text{ cost}}  \\
    \text{perturbed velocity:} \quad &
        \left\{
        \begin{aligned}
                \bm{u}(\bm{x}+h\bm{e}_i) &= (\bm S+ \Delta\bm S^{(i)})(\bm f+ \Delta\bm f^{(i)}) \\
                &=  \underbrace{\bm u(\bm x)}_{\text{compute once}}+ \underbrace{\Delta\bm S^{(i)} \bm f^{(i)}+   \bm S\Delta \bm f^{(i)}+ \Delta\bm S^{(i)}\Delta\bm f^{(i)} }_{\mathcal{O}(\epsilon^{-1})\text{ cost}}         
        \end{aligned}
        \right.
        \label{ea11c} 
\end{align}

\begin{figure}[H]
	\begin{subfigure}{0.45\textwidth}
		\centering
		\includegraphics[width=\textwidth,height=2.2 in]{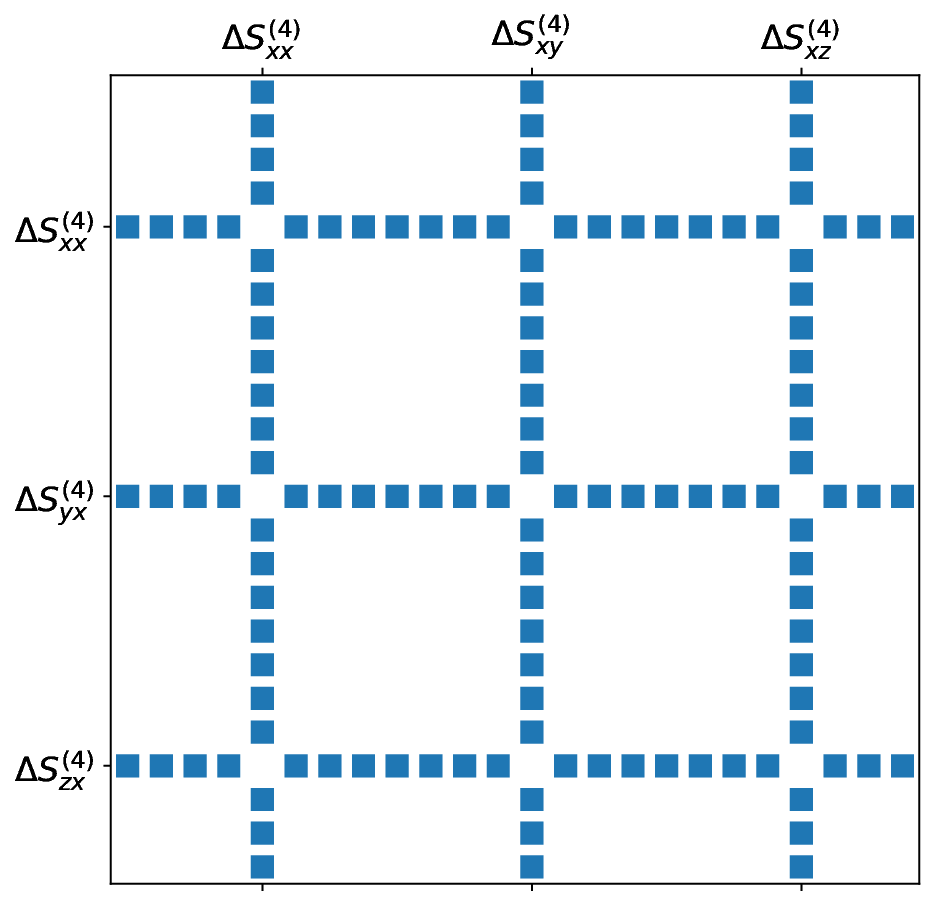}
		\caption{Non-zero entries in perturbed Stokeslet}
		\label{DS}
	\end{subfigure}
	\hfill
	\begin{subfigure}{0.45\textwidth}
		\centering
		\includegraphics[height=2.2 in]{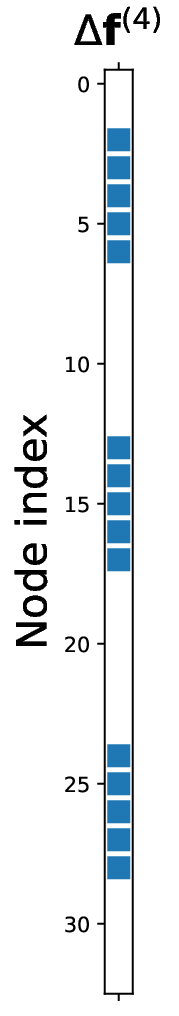}
		\caption{Non-zero entries in perturbed forces}
		\label{Df}
	\end{subfigure}
	\caption{(a) Non-zero entries in perturbed Stokeslet $\Delta \bm{S}^{(4)}$ due to a perturbation in the fourth node, and (b) Non-zero entries in perturbed forces $\Delta \bm{f}^{(4)}$ due to a perturbation in the fourth node. Note that for our model problem we only have two non-zero rows and columns, rather than three, due to a planar motion (\textit{i.e.}, $z=0$). }
\end{figure}

The matrices $\Delta \bm{S}^{(i)}$ have only three non-zero rows and columns (see \cref{DS}). Therefore, the cost of matrix-vector multiplications involving the perturbed Stokeslet matrix $\Delta \bm{S}^{(i)}$ and forces  $\Delta \bm{f}^{(i)}$ in equation \eqref{ea11c} can be reduced from $\mathcal{O}(\epsilon^{-2})$ for a dense matrix to $\mathcal{O}(\epsilon^{-1})$. Furthermore, due to the second derivatives involved in the bending forces (equation \ref{e42}), perturbing a node changes at most 15 entries of forces (Figure \ref{Df}). Specifically, perturbed forces change 15 entries for an interior node, 12 for a node next to the boundaries (indices 2 and $N-1$), and 9 for the boundary nodes (indices 1 and $N$). Consequently, due to this sparsity, the matrix-vector product involving a perturbed expression (e.g., $\bm{S} \Delta\bm{f}^{(i)}$) costs $\mathcal{O}(\epsilon^{-1})$. Thus, the total computational cost for computing all the columns of the Jacobian with this approach is $\mathcal{O}(\epsilon^{-2})$, as opposed to $\mathcal{O}(\epsilon^{-3})$ for naive approach.

\begin{figure}[H]
     \centering
     \includegraphics[width=0.7\linewidth,height=2.5in]{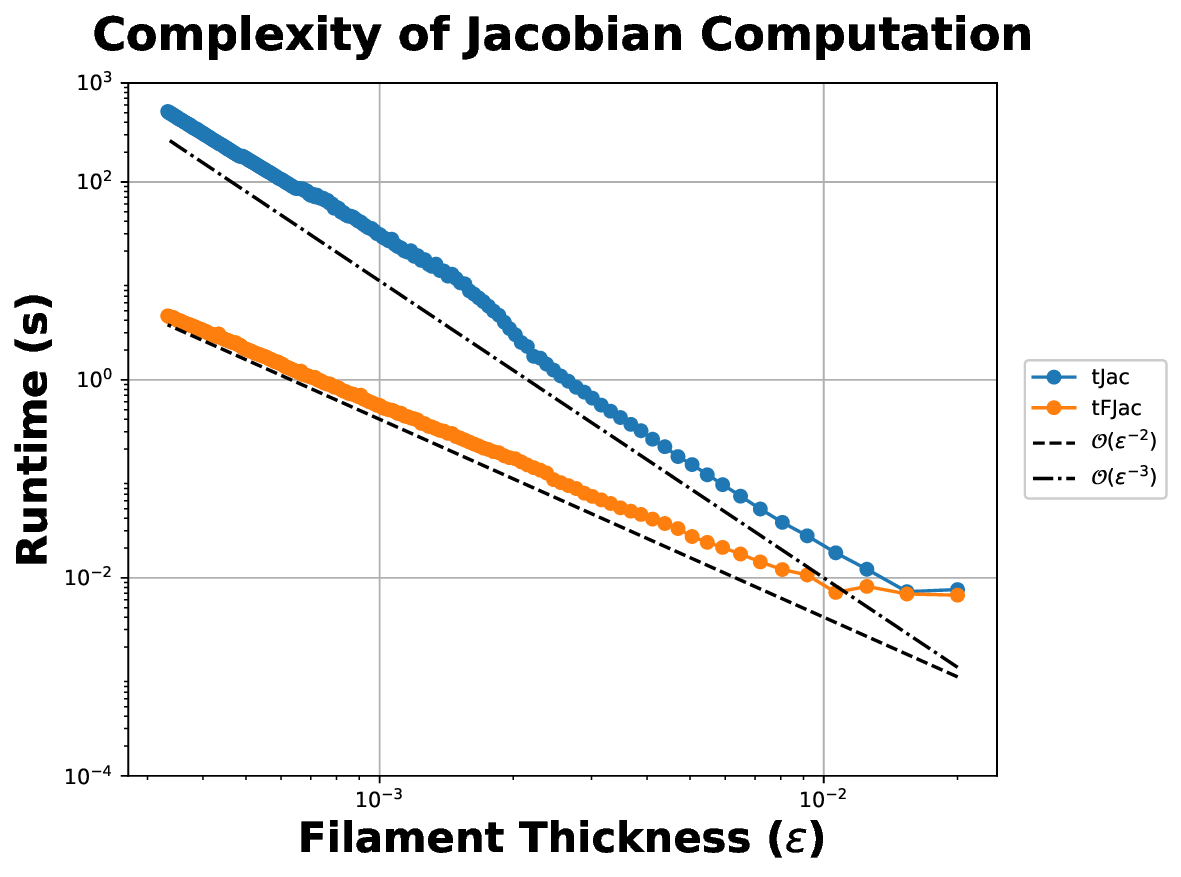}
     \caption{Computational complexity of Jacobian computation using naive and fast versions of FD approximation considering varying thickness $\epsilon$ and $S_C=0.05$.}
     \label{fig_jac_o}
 \end{figure}

We validate the complexity of the fast version of Jacobian computation in Figure \ref{fig_jac_o}, where we observe that the asymptotic runtimes of the fast version scales as $\mathcal{O}(\epsilon^{-2})$, as opposed to $\mathcal{O}(\epsilon^{-3})$ for the naive implementation.

\begin{figure}[H]
    \begin{subfigure}{0.45\textwidth}
         \includegraphics[width=\textwidth,height=2.2in]{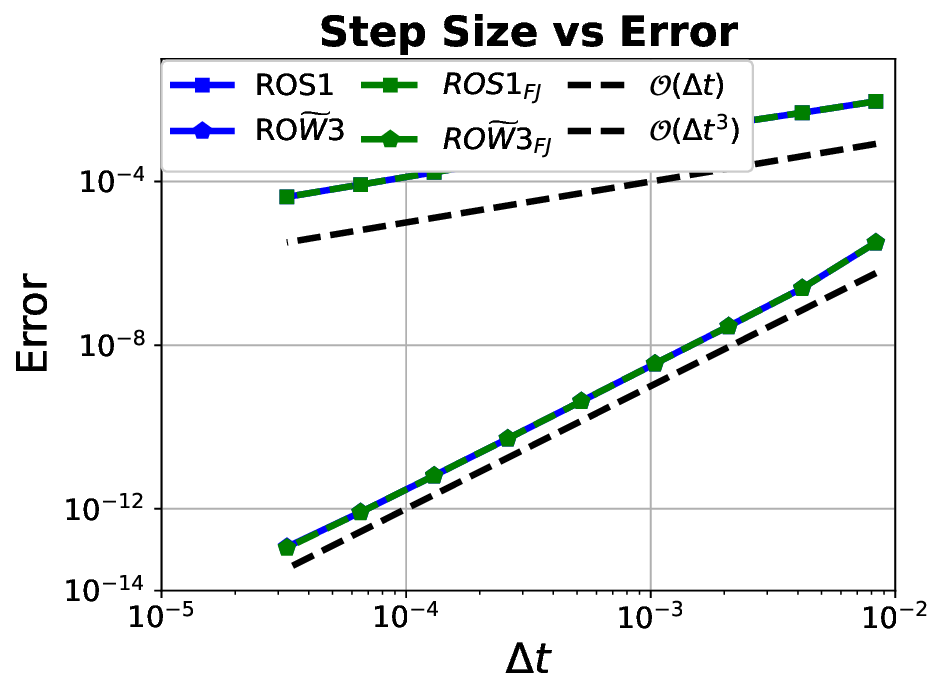}
         \caption{Convergence plots (\textbf{base case})}
         \label{fig20Ca}
     \end{subfigure}
     \hfill
     \begin{subfigure}{0.45\textwidth}
         \includegraphics[width=\textwidth,height=2.2in]{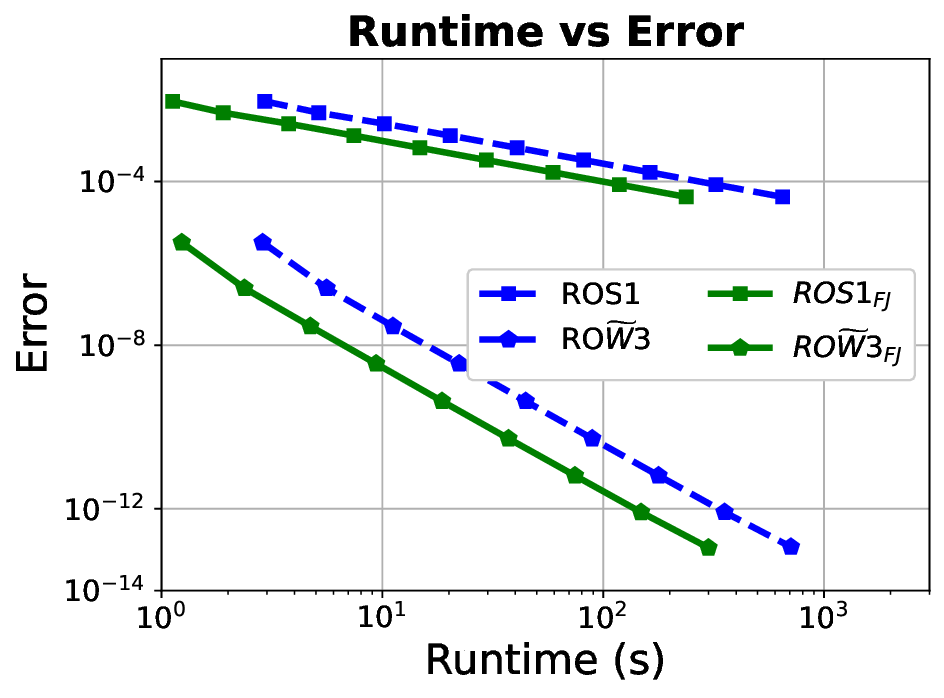}
        \caption{Runtime plots}
         \label{fig20Cb}
     \end{subfigure}
     \hfill
         \begin{subfigure}{0.45\textwidth}
         \includegraphics[width=\textwidth,height=2.2in]{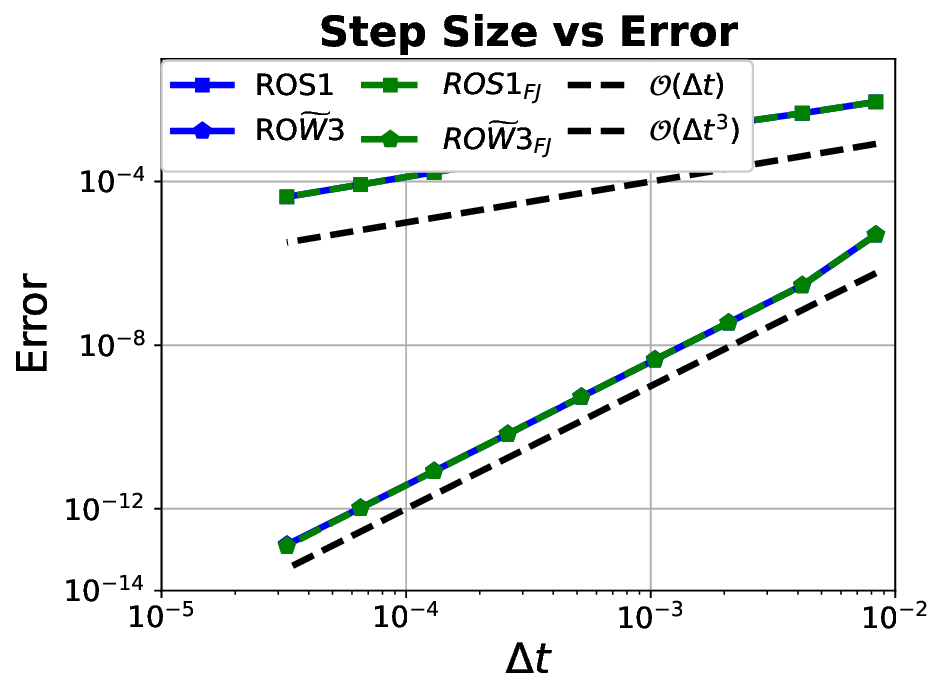}
         \caption{Convergence plots (\textbf{slender filament})}
         \label{fig20Cc}
     \end{subfigure}
     \hfill
     \begin{subfigure}{0.45\textwidth}
         \includegraphics[width=\textwidth,height=2.2in]{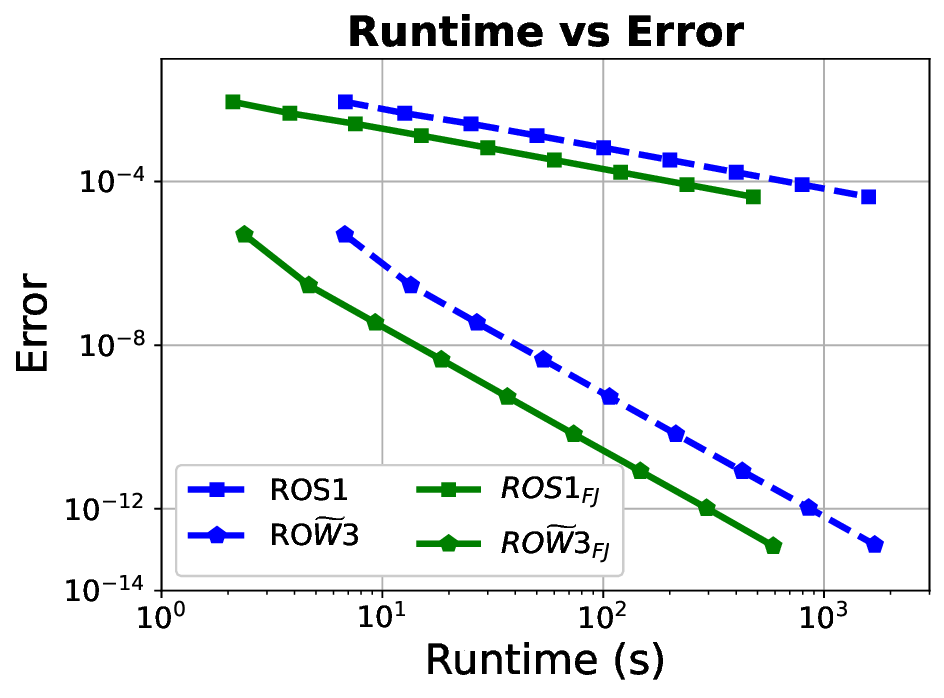}
        \caption{Runtime plots}
         \label{fig20Cd}
     \end{subfigure}
     \caption{Convergence and runtime plots using fast and naive approaches for computing FD Jacobians. \textbf{FJ} in the legends indicates the fast Jacobian computation approach. (a) Convergence plots for the base case, (b) runtime plots for the base case, (c) convergence plots for the slender filament, and (d) runtime plots for the slender filament.}
     \label{fig20C}
\end{figure}

In Figure \ref{fig20C}, we show convergence and runtime plots for ROS1 and \text{RO$\widetilde{W}$3} with the Jacobian computed using naive and fast implementations of FD approximation. We consider the base $(S_C,\epsilon)=(0.05,0.01)$ and slender filament $(S_C,\epsilon)=(0.05,0.0071)$ cases. We do not consider the stiff cases, since the speedup using the fast Jacobian construction depends on $\epsilon$ and is independent of $S_C$.

    \textbf{Convergence.}  The convergence plots \ref{fig20Ca} and \ref{fig20Cc} indicate that both approaches for FD-based Jacobian computations lead to identical accuracy and convergence for both ROS1 and \text{RO$\widetilde{W}$3}. This indicates that rewriting the Jacobian FD column approximation with sparse matrices does not introduce any significant rounding errors.
  
    \textbf{Efficiency.} From the runtime plot (Figure \ref{fig20Cb}), we observed an average performance gain of about $2.74$ and $2.38$ times when using the fast FD-based Jacobian computation compared to the naive version, for ROS1 and \text{RO$\widetilde{W}3$}, respectively.

    As we decrease $\epsilon$ to $0.0071$ (in Figure \ref{fig20Cd}, we observed an average speedup of about $3.33$ and $2.89$, for ROS1 and \text{RO$\widetilde{W}3$}, respectively. This indicates that the speedup gain from the fast Jacobian implementation will increase as we decrease $\epsilon$. These speedups agree closely with the expected scaling $\mathcal{O}(\epsilon)$. Specifically, we observe a speedup of about $1.2$ times when we decrease $\epsilon$ by a factor of $1.4$, close to the expected $1.4$ times speedup.

 \subsubsection{ROS Methods Using Automatic Differentiation Jacobians}\label{ss73}

 Recent advances in automatic differentiation (AD) have made it a viable strategy to efficiently construct Jacobians for ROS methods \citep{steinebach_construction_2023}. Here we compare the accuracy and efficiency of AD-based Jacobians against FD Jacobians using the model problem from Section \ref{s4}. In particular, we are interested in two questions: (i) how does the computational complexity of AD compare to FD, (ii) does AD's immunity to round-off errors prevent order reduction for ROS methods that require exact Jacobians? To answer these questions, we consider three third-order methods, ROS3, ROW3, \text{RO$\widetilde{W}$3}, and compute all AD Jacobians using the \texttt{jacobian} function from JAX \citep{jax2018github}.

\begin{figure}[H]
     \centering
     \includegraphics[width=0.7\linewidth,height=2.2in]{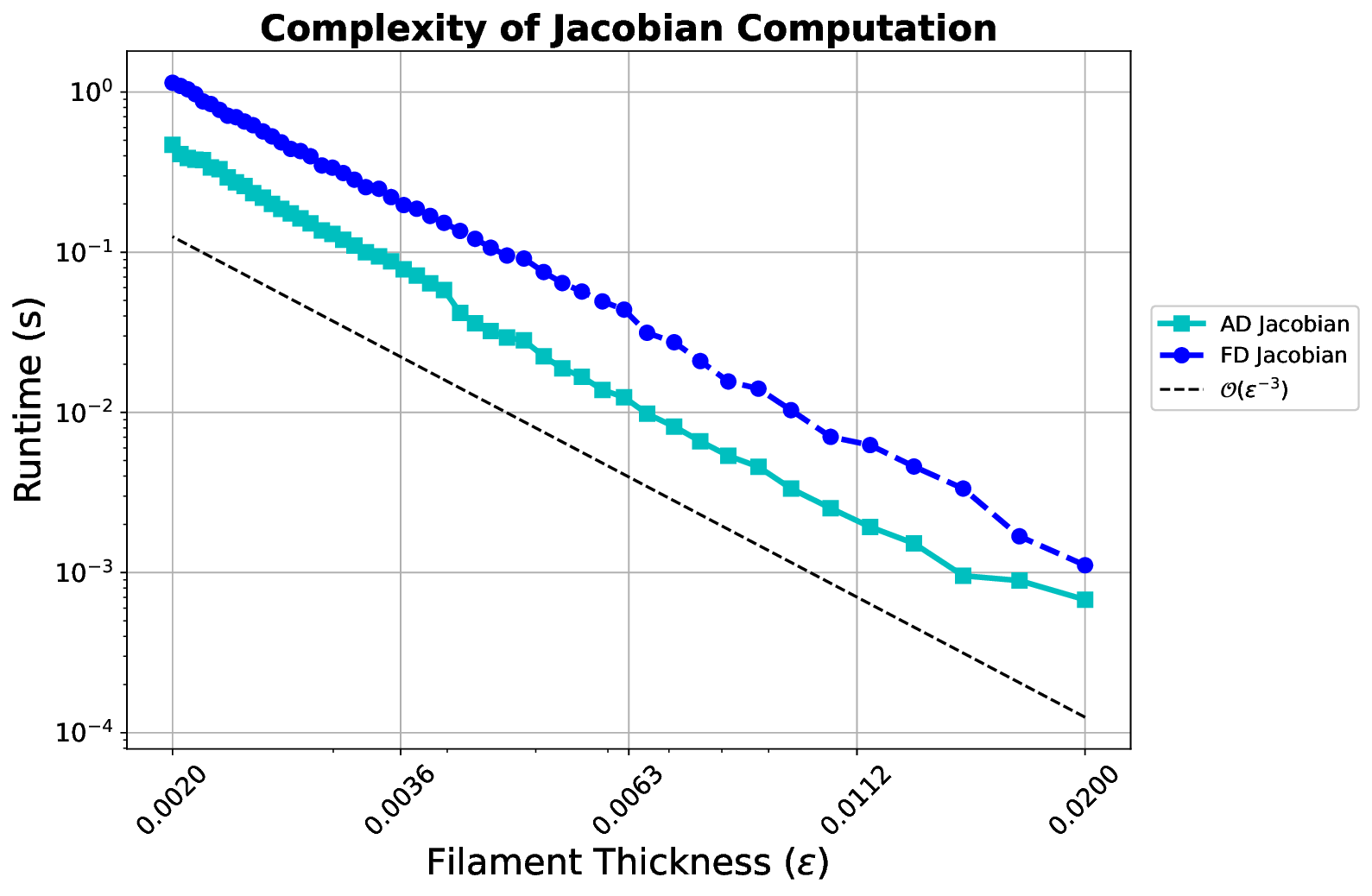}
     \caption{Computational complexity of Jacobian computation using AD and FD.}
     \label{fig_jac_AD}
 \end{figure}

 \textbf{Computational Complexity.} In Figure \ref{fig_jac_AD},  we plot the run times of the AD and FD Jacobian construction as a function of $\epsilon$. We observe that AD computation times are faster; however, both algorithms have an asymptotic runtime of $\mathcal{O}(\epsilon^{-3})$. These simulations run in JAX; therefore, the timing curve for the FD Jacobian is not identical to the one shown in Figure \ref{fig_jac_o}, which was run in MATLAB. Lastly, we note that the computational complexity is independent of $S_C$, which simply scales a portion of the right-hand-side $\bm{u}$.

\begin{figure}[H]
	\begin{subfigure}{0.48\textwidth}
		\includegraphics[width=\textwidth]{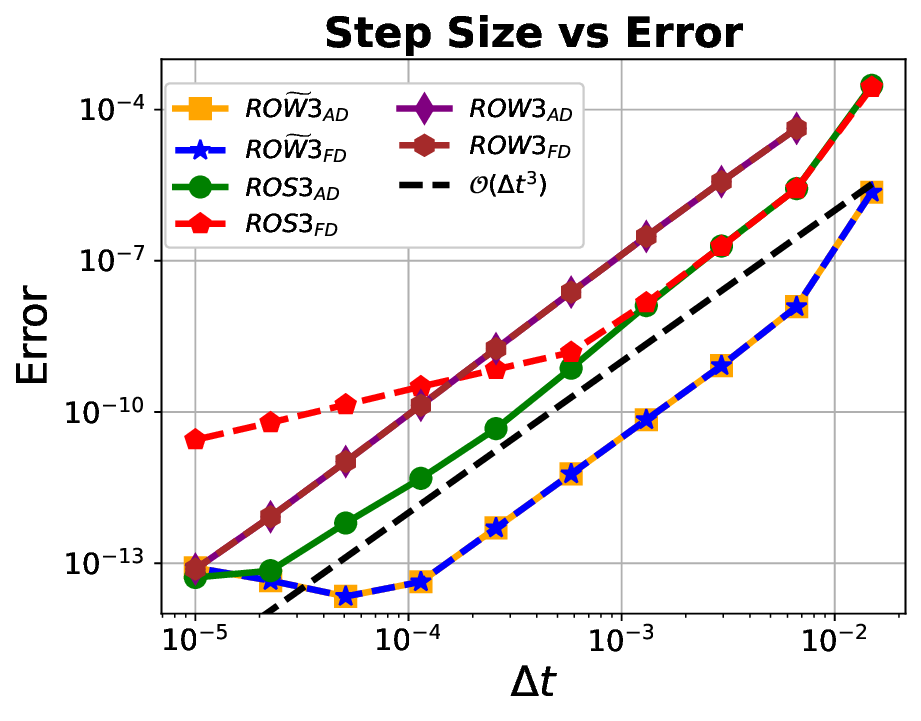}
		\caption{Convergence plots}
		\label{fig21a}
	\end{subfigure}
	\hfill
	\begin{subfigure}{0.48\textwidth}
		\includegraphics[width=\textwidth]{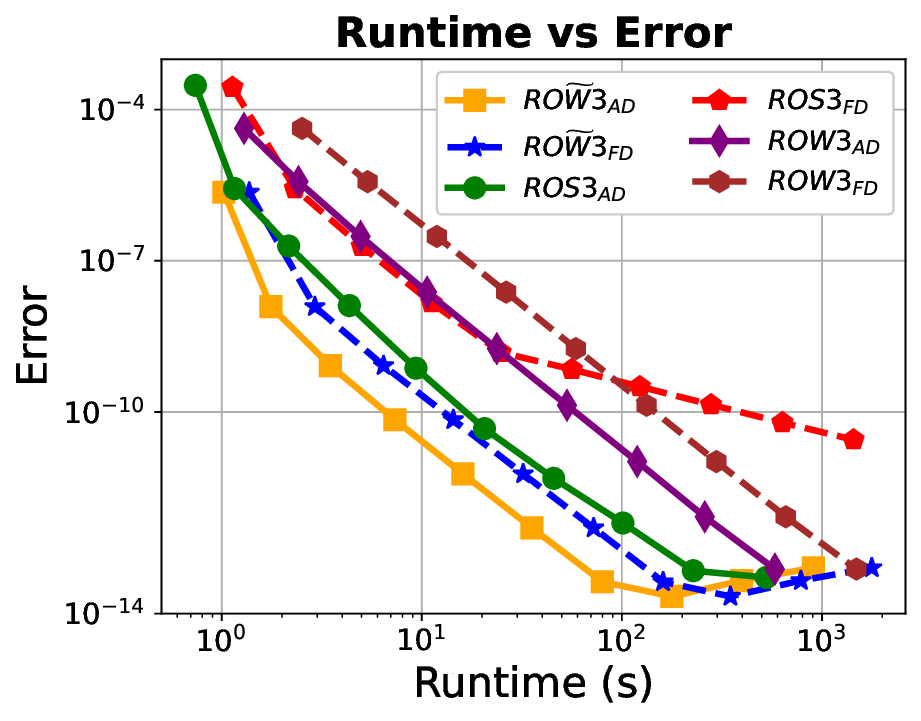}
		\caption{Runtime Plots}
		\label{fig21b}
	\end{subfigure}
	\caption{ (a) Convergence and (b) runtime plots using AD (dashed lines) and FD (solid lines) for computing the Jacobian for the base case $(S_C,\epsilon)=(0.05,0.01)$.}
	\label{fig21}
\end{figure}

 \textbf{Convergence.} The convergence plot (Figure \ref{fig21a}) shows that AD and FD provide identical precision for all methods, except for ROS3 that requires an exact Jacobian. ROS3 experiences a reduction in expected order of convergence when FD is utilized for Jacobian computation, starting at around $\Delta t=2\times 10^{-4}$. This occurs due to the limited accuracy of the FD approximations \eqref{e47} (note that reducing $h$ further will introduce round-off error that prevents further improvement in accuracy). Although higher-order finite differences can mitigate this round-off, they will also increase computational complexity significantly.  In contrast, $W$ and $\widetilde{W}$ methods are robust to errors in FD Jacobians without increasing computational cost.

 \textbf{Efficiency.} Since asymptotic run times for AD and FD are identical in both $\epsilon$ and $S_C$, we only need to run the base case simulation $(S_C,\epsilon)=(0.05,\epsilon)$ to investigate efficiency. Figure \ref{fig21b} shows runtime plots. As expected from the asymptotic runtime, AD-based $W$ and $\widetilde{W}$ methods (solid lines) are approximately two times faster than the corresponding FD-based counterparts (dashed lines). For ROS3, the improved accuracy of AD leads to more than two times speedup for errors less than $10^{-8}$. Specifically, this speedup rises to about 40 times for around 10-digit accuracy.

\subsubsection{ROW Methods with a Time Lagged Jacobian (TLJ)}

We now explore another strategy to reduce total runtime by reusing the Jacobian over several consecutive time steps; this is typically referred to as a time-lagged Jacobian (TLJ) \citep{verwer1980rosenbrock}. TLJ reduces computational cost by avoiding Jacobian recomputation at each time step, and is especially applicable for $W$ methods that retain order of accuracy even when the Jacobian is not exact.  In this section, we explore how TLJ affects the accuracy and stability of the $ W$-methods ROW1 and ROW3.

Before beginning our discussion, we define the following variables.
\begin{center}
    \begin{tabular}{|c|c|}
    \hline
       Variable  & Description\\
         \hline
       $C_J$   & Runtime for Jacobian computation\\
                  \hline
       $C_O$   & Runtime excluding Jacobian computation\\
                  \hline
       $T_{\text{ROW}}$  & Runtime for a ROW method\\
       \hline
       $T_{\text{ROW-TLJ}}$& Runtime for a ROW method using TLJ\\
         \hline
                $S_{\text{TLJ}}$   & Speedup using TLJ; $S_{\text{TLJ}}=\frac{T_{\text{ROW}}}{T_{\text{ROW-TLJ}}}$\\
                  \hline
       $k$  & TLJ skip rate, which must be a positive integer\\
       \hline
       $t_f$  & Final time\\
       \hline
       $N_{\text{t}}$& Total time steps to reach final time $t_f$\\
         \hline
        $\Delta t$   & Step size; $\Delta t=\frac{t_f}{N_t} $\\
                  \hline
       $\Delta t_{\text{Jmax}}$   & user-selected threshold step size below which Jacobian variation is small\\
                  \hline
    \end{tabular} 
\end{center}

\textbf{Asymptotic Speedup Using TLJ. } 
If we only compute the Jacobian at time steps indexed by $1, k+1, 2k+1, \ldots$, then the runtimes  for ROW and ROW-TLJ, $T_{\text{ROW}}$ and $T_{\text{ROW-TLJ}}$, and speedup using TLJ, $ S_{\text{TLJ}} $, are given by: 
\begin{align}
T_{\text{ROW}}&=N_t(C_J + C_O),&T_{\text{ROW-TLJ}}&=N_t(\frac{C_J}{k} + C_O),&S_{\text{TLJ}} &= \frac{k(C_J + C_O)}{C_J + k C_O}.\label{e255}
\end{align} 
Based on the analysis in Section \ref{s6}, we can approximate  $C_J$ and $C_O$  for our model problem as: $C_J = c_1 \epsilon ^{-2.5}, C_O = c_2 \epsilon^{-2}$, where $c_1,c_2$ are arbitrary constants that depend on time-integrators. Here, the exponent in the Jacobian cost is $-2.5$, instead of $-3$ based on the best-fit line for sufficiently large $\epsilon$ (specifically, we fitted a line using $\epsilon$ up to $1e-2$, as opposed to $3.33e-04$ in figure \ref{fig_jac_o}). Using these, the speedup $S_{\text{TLJ}}$ using a TLJ method is given by:
\begin{align}
    S_{\text{TLJ}} =\frac{  1 + r\sqrt{\epsilon} }{\frac{1}{k} + r\sqrt{\epsilon}}, \quad \text{where} \quad
    \renewcommand\arraystretch{1.25}
    \begin{tabular}{c|cc}
          & ROW1 & ROW3 \\ \hline
        $r=\frac{c_2}{c_1 }$& 0.05   & 0.23 
    \end{tabular}.\label{e25}
\end{align} 
\textbf{Selecting TLJ Skip Rate $k$.}
We apply a simple strategy in which the Jacobian is updated on a user-selected timescale $\Delta t_{\text{Jmax}}$. This stepsize should be chosen to avoid allowing a stale Jacobian to compromise stability or accuracy of a ROW method. In practice, $\Delta t_{\text{Jmax}}$ depends on the system being solved and on the robustness of the method to inexact Jacobians. To ensure that $k\Delta t < \Delta t_{\text{Jmax}}$ we require that
\begin{align}\label{e28}
    k=\bigg\lceil\frac{\Delta t_{\text{Jmax}}N_t }{t_f}\bigg\rceil,
\end{align}
where $ \lceil\cdot\rceil$ is the ceiling function.

We chose $\Delta t_{\text{Jmax}}$ to be roughly the stepsize at which each method begins showing convergence at its expected order of accuracy. From the convergence plots in subsection \eqref{ss71}, we observed that ROS1 (first-order method) starts to achieve first-order convergence at $\Delta t = \frac{1}{100}$. In contrast, for ROW3 (third-order method), we needed a slightly smaller value to achieve third-order convergence at small time steps. For this model problem, the chosen values for $\Delta t_{\text{Jmax}}$ are:
\begin{align}
    \begin{aligned}
    \text{ROW1:} \quad \Delta t_{\text{Jmax}}=\frac{1}{100}, &&
    \text{ROW3:} \quad \Delta t_{\text{Jmax}}=\frac{1}{400}.
    \end{aligned}
\end{align}

\begin{figure}[H]
	\begin{subfigure}{0.48\textwidth}
		\includegraphics[width=\textwidth]{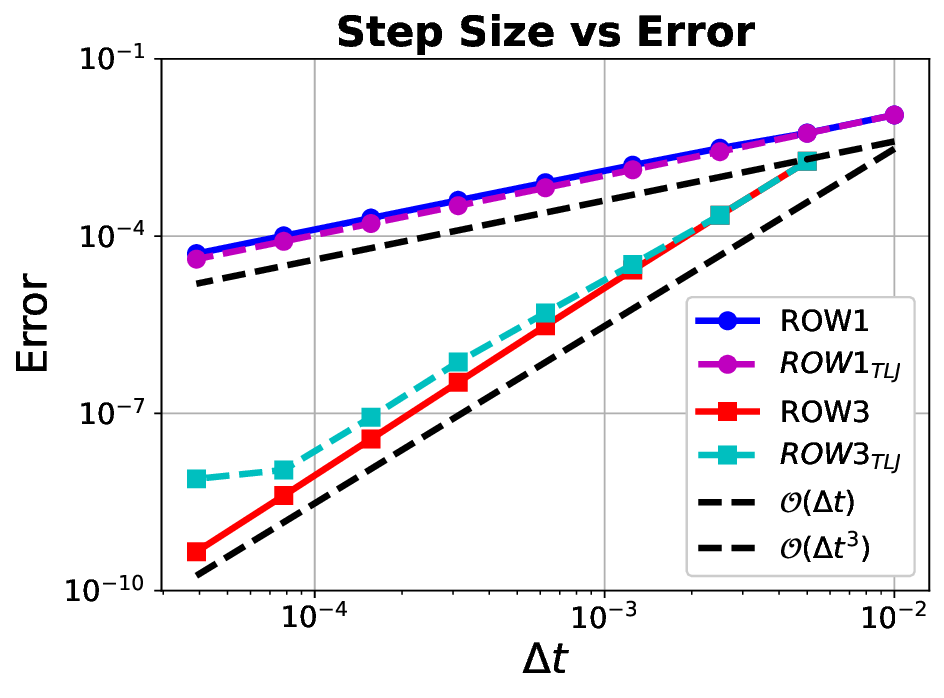}
		\caption{Convergence Plots (\textbf{Base case})}
		\label{fig_tlja}
	\end{subfigure}
	\hfill
	\begin{subfigure}{0.48\textwidth}
		\includegraphics[width=\textwidth]{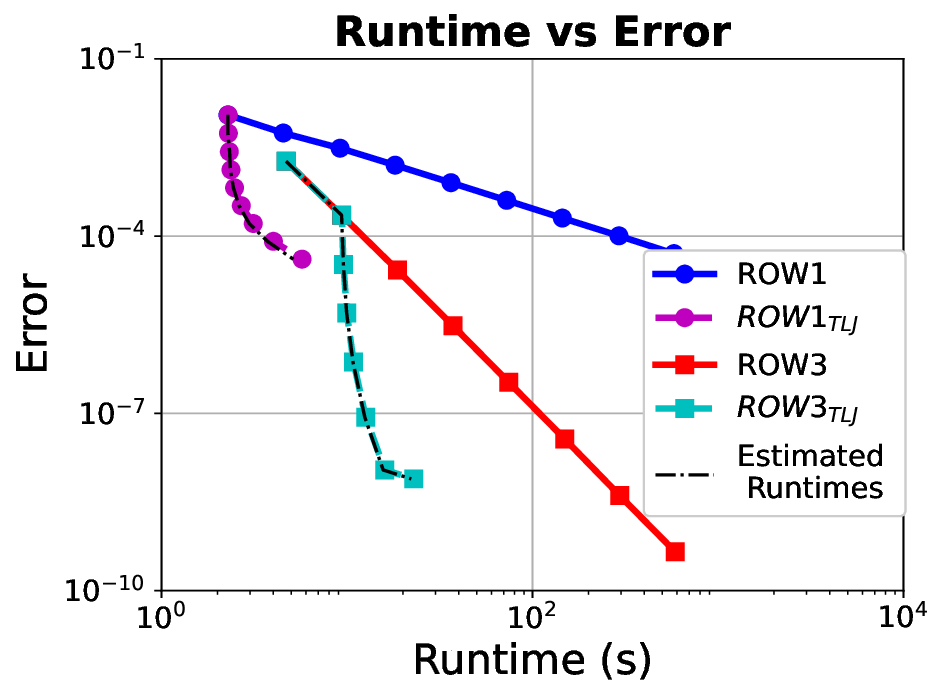}
		\caption{Runtime Plots }
		\label{fig_tljb}
	\end{subfigure}

	\begin{subfigure}{0.48\textwidth}
		\includegraphics[width=\textwidth]{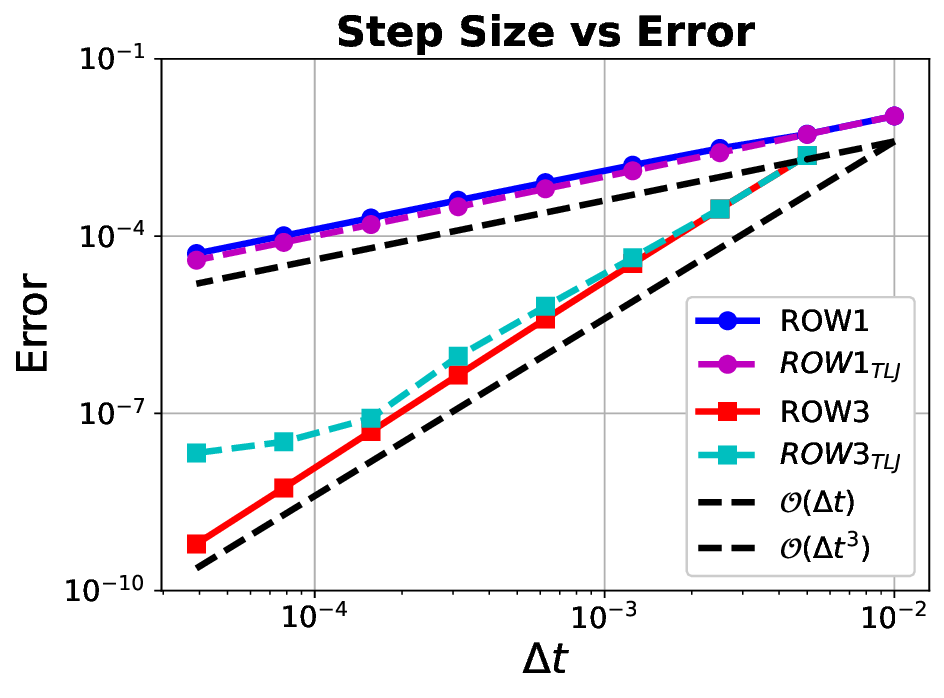}
		\caption{Convergence Plots (\textbf{Slender filament})}
		\label{fig_tlje}
	\end{subfigure}
	\hfill
	\begin{subfigure}{0.48\textwidth}
		\includegraphics[width=\textwidth]{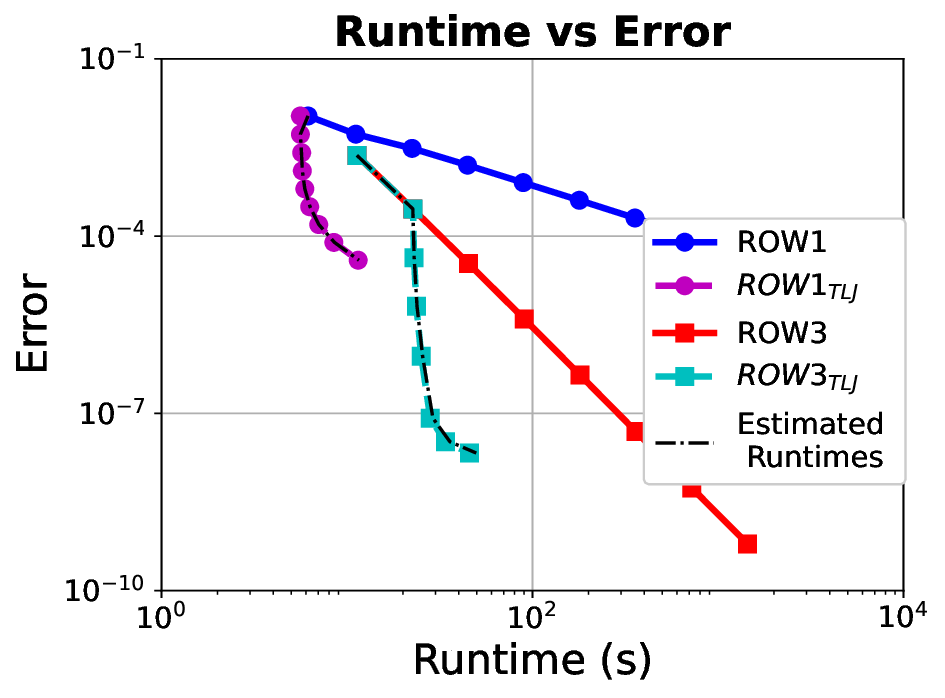}
		\caption{Runtime Plots }
		\label{fig_tljf}
	\end{subfigure}
	\caption{Convergence and runtime plots using time-lagged Jacobian (TLJ) for ROW1 and ROW3. (a) Convergence plots for the base case, (b) runtime plots for the base case,  (c) Convergence plots for the slender filament, and (d) runtime plots for the slender filament.}
	\label{fig_tlj2}
\end{figure}

 In Figure \ref{fig_tlj2}, we visualized the convergence and runtime plots for ROW1 and ROW3 with and without TLJ. In both cases, the Jacobian computation is done with the naive FD approximation. As the saving formula \eqref{e25} indicates, the speedup using TLJ is independent of $S_C$; therefore, we only run simulations for the base and slender-filament cases.
 
\textbf{Convergence.}
Figures \ref{fig_tlja} and \ref{fig_tlje} show convergence plots for ROW1 and ROW3 for the base and slender-filament cases. 
The convergence plots show that skipping the Jacobian has little effect on the solution accuracy for ROW1. However, time-lagging for ROW3 shows a slight reduction in convergence order at small step sizes (Figure \ref{fig_tlje}). 

          \textbf{Efficiency.}
            Figures \ref{fig_tljb} and \ref{fig_tljf} show efficiency plots for ROW1 and ROW3 for the base and slender filament.  The runtime plot for the base case (Figure \ref{fig_tljb}) shows that TLJ methods run faster than methods that compute the Jacobian at each time step. Specifically, skipping the Jacobian computation yields mean speedups of approximately 30 and 8 times for ROW1 and ROW3, respectively.
            Runtime plots for the slender filament case (Figure \ref{fig_tljf}) indicate mean speedups of about 35 and 10 times for ROW1 and ROW3, respectively. This shows that speedup increases as $\epsilon$ decreases, consistent with the estimated savings from the TLJ method reported in \eqref{e25}. In both cases, the estimated runtimes (dashed-dotted lines approximated using \ref{e25}) for each TLJ method are almost identical to the actual runtimes.

\subsection{Combining Enhancement Strategies}

Here we combine the most effective strategies discussed in Subsection \ref{SS72} to further reduce the computational time of ROW1 and ROW3 (the two $W$-methods from Table \ref{tabm}). To simulate a scenario in which automatic differentiation is unavailable, we pair a time-lagged Jacobian (TLJ) with fast finite-difference Jacobian computation. Conversely, we also pair a time-lagged Jacobian with AD Jacobian computation. Lastly, we include $RO\tilde{W}3$ without time lagging, and using either fast finite-difference or AD-based Jacobian construction. We show that, even without time lagging, the small error constant of $RO\tilde{W}3$ makes it the most competitive method when high accuracy is required.

\begin{figure}[H]
	\includegraphics[width=\textwidth,height=2.2in]{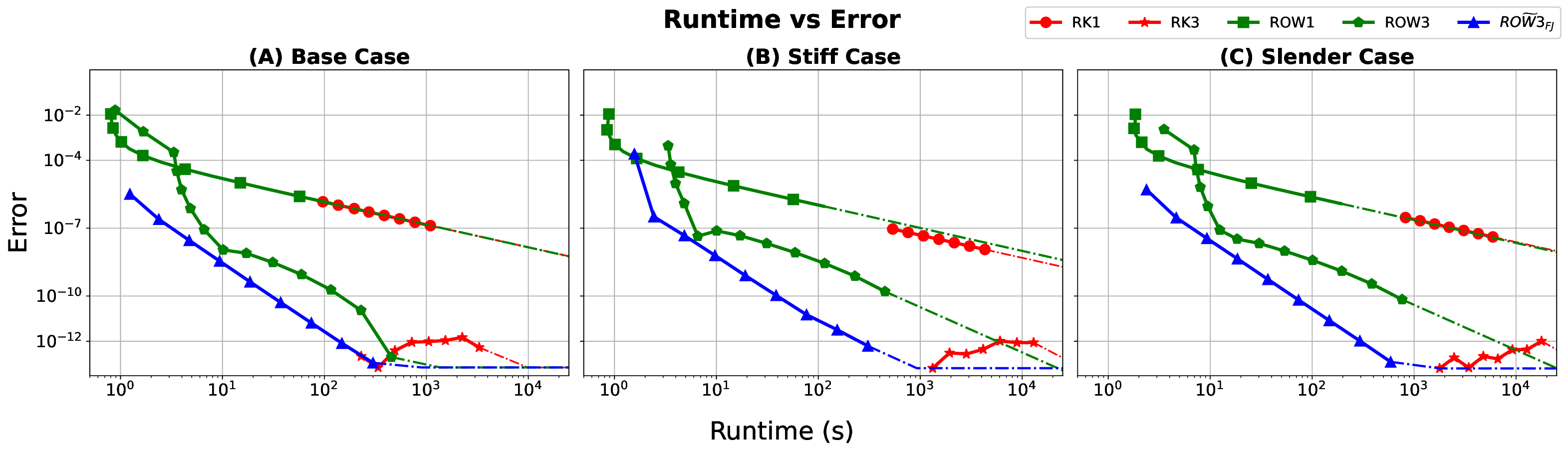}
	\caption{Runtime plots for: (\textbf{A}) base, (\textbf{B}) stiff, and (\textbf{C}) slender filament cases using ROW1, ROW3, RK1, and RK3. For ROW methods, we combined two enhancement strategies, namely: TLJ and Fast Jacobian computation. Dashed lines represent extrapolated errors.}
	\label{fig_FJ_TLJ}
\end{figure}

\begin{figure}[H]
	\includegraphics[width=\textwidth,height=2.2in]{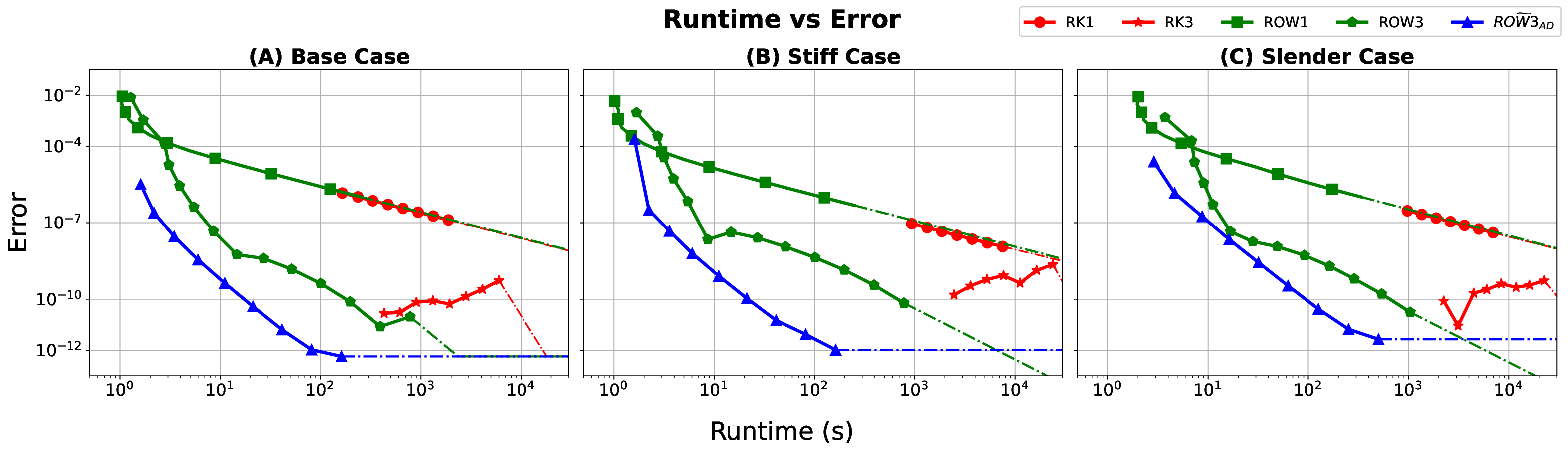}
	\caption{Runtime plots for: (\textbf{A}) base, (\textbf{B}) stiff, and (\textbf{C}) slender filament cases using ROW1, ROW3, RK1, and RK3. For ROW methods, we combined two enhancement strategies, namely: TLJ and AD Jacobian computation. Dashed lines represent extrapolated errors.}
	\label{fig_AD_TLJ}
\end{figure}

\textbf{Efficiency with Time-Lagged, Fast FD Jacobians.}
Figures \ref{fig_FJ_TLJ}\textbf{(A)}, \ref{fig_FJ_TLJ}\textbf{(B)}, and \ref{fig_FJ_TLJ}\textbf{(C)} show work-precision plots for the base, stiff, and slender filaments, respectively. Combining a time-lagged Jacobian (TLJ) with fast Jacobian computation enables ROW1 to achieve approximately 2--4 digits of accuracy in one second or less for the base and stiff cases [Figures \ref{fig_FJ_TLJ}\textbf{(A)} and \ref{fig_FJ_TLJ}\textbf{(B)}] and in approximately two seconds for the slender case [Figure \ref{fig_FJ_TLJ}\textbf{(C)}]. Even at these coarse accuracy levels, the ROW1 solutions are visually indistinguishable from the reference solution and are fully suitable for practical biofluid simulations. Relative to the first stable RK1 solution, ROW1 provides a speedup factor of approximately 300--2000. For higher accuracy, $\text{RO}\widetilde{W}3$ with the fast Jacobian approach is preferred; although it cannot use a TLJ because it is not a $W$-method, it still achieves approximately 7-digit accuracy with a speedup factor of roughly 250--500 over RK1.

\textbf{Efficiency with Time-Lagged, AD Jacobians.}
Figures \ref{fig_AD_TLJ}\textbf{(A)}, \ref{fig_AD_TLJ}\textbf{(B)}, and \ref{fig_AD_TLJ}\textbf{(C)} show work-precision plots for the base, stiff, and slender filaments, respectively. Combining a time-lagged Jacobian (TLJ) with automatic-differentiation (AD) Jacobian computation enables ROW1 to achieve approximately 2--4 digits of accuracy in about one second for the base and stiff cases [Figures \ref{fig_AD_TLJ}\textbf{(A)} and \ref{fig_AD_TLJ}\textbf{(B)}] and in approximately two seconds for the slender case [Figure \ref{fig_AD_TLJ}\textbf{(C)}]. At these accuracy levels, the ROW1 solutions are visually indistinguishable from the reference solution in the eyeball norm. Relative to the first stable RK1 solution, ROW1 provides a speedup factor of approximately 150--1000. At higher accuracy, $\text{RO}\widetilde{W}3$ with an AD-computed Jacobian is preferred, providing roughly 7-digit accuracy and a speedup factor of approximately 150--850 over RK1.

 \section{Conclusion}\label{s8}

Using Rosenbrock (ROS) time-stepping method, we reduce simulation time by 20 to 2000 times compared to explicit RK (ERK) methods for a filament-based biofluid model that uses the method of regularized Stokeslets. To understand this speedup, we studied how it related to key model parameters and investigated the factors that influence the system's numerical stiffness. Furthermore, we provided approximate scaling laws to quantify the speedups achieved by ROS methods compared with ERK methods. Additionally, we discussed several strategies to speedup simulations beyond the naive implementations of the ROS methods considered. Among the ROS methods considered, we found that the third-order method, \text{RO$\widetilde{W}$3}, yields a more precise solution at a cost not significantly higher than that of other methods. In terms of stability, first-order ROS1 is slightly more stable than other ROS methods considered, which means we can use larger step sizes with it.
Overall, the findings of the current study encourage us to test ROS-type integrators across a broader range of related biofluid applications, with the aim of significantly reducing simulation times and enhance understanding of this field.

 \textbf{Availability of Codes.} All the codes used will be made public upon publication; currently they are available in a Github repository \cite{GIT_REP}.
 \appendix

\section{Non-Autonomous Form of Time Integrators and Coefficients}\label{app1}

An RK method \citep{ wanner1996} in non-autonomous form can be written as:
\begin{subequations}
    \begin{align}
 \bm{K_i}&=\Delta t \bm{u}(t_n+c_i \Delta t,\bm{x}_n+ \sum_{j=1}^{s} a_{ij} \bm{K_i}),\quad i=1,\ldots,s,\\
 	\bm{x}_{n+1}&= \bm{x}_n+\sum_{i=1}^{s} b_i \bm{K_i}. 
\end{align}
\end{subequations}
A ROS method \citep{ wanner1996} in non-autonomous form can be written as:
\begin{subequations}\label{eA2}
    \begin{align}
    \bigg(I - \Delta t\gamma_{ii}\bm{J}_{\bm{x}}\bigg)\bm{K_i} &= \Delta t\bigg( \bm{u}(t_n+\Delta t\alpha_i,\bm{x}_n +
	\sum_{j=1}^{i-1} \alpha_{ij}\bm{K_j} )+ \bm{J}_{\bm{x}}
	\sum_{j=1}^{i-1}\gamma_{ij} \bm{K_j} +\gamma_i\Delta t \bm{J}_t\bigg)  ,\\ i = 1, \ldots , s, \notag\\
	\bm{x}_{n+1} &= \bm{x}_n +
	\sum_{i=1}^{s} b_i \bm{K_i},
\end{align}
\end{subequations}
where $\alpha_i= \sum_{j=1}^{i-1} \alpha_{ij},\gamma_i= \sum_{j=1}^{i} \gamma_{ij} $. Usually, all $\gamma_{ii}$'s are taken as a scalar $\gamma$ (including ROS, $W$, and $\widetilde{W}$ methods listed in table \ref{tabm}). Moreover, $\bm{J}_{\bm{x}}=\frac{\partial \bm{u}}{\partial \bm{x}}(t_n,\bm{x}_n),\bm{J}_t=\frac{\partial \bm{u}}{\partial t}(t_n,\bm{x}_n)$. 

Any non-autonomous ODE system can be recast as an autonomous system by introducing an additional variable for time, $\bm{x}_{N+1}=t$, and augmenting the corresponding ODE  $\dot{\bm{x}}_{N+1}=1$ to the system \citep{wanner1996}.

The matrix-vector multiplication involved in \eqref{eA2} can be avoided with a suitable transformation of the coefficients, which allows us to rewrite \eqref{eA2} as follows \citep{wanner1996, Lang01}:
\begin{subequations}\label{eA5}
\begin{align}
	\left(\frac{I}{\Delta t \gamma_{ii}} -\bm{J}_{\bm{x}}\right) \bm{S_i} &= \bm{u}\left(t_n+\Delta t\alpha_i  , \bm{x}_n+\sum_{j=1}^{i-1} a_{i j} \bm{S_j}\right) 
	+\sum_{j=1}^{i-1}\left(\frac{c_{i j}}{\Delta t}\right) \bm{S_j}+\gamma_i \Delta t \bm{J}_t,\label{e21} \\
	\bm{x}_{n+1}  &= \bm{x}_n +
	\sum_{i=1}^{s} m_i \bm{S_i},\label{e22}
\end{align}
\end{subequations}
where $\bm{S_i} =\displaystyle \sum_{j=1}^i \gamma_{ij} \bm{K_j}$. With $\Gamma=\left(\gamma_{i j}\right)_{i, j=1}^s$ and $\gamma_{i i} \neq 0$ for all $i$, the coefficients can be obtained using the following transformations \citep{wanner1996}: 
\begin{subequations}\label{eA3}
\begin{align}
	\left(a_{i j}\right)_{i, j=1}^s= & \left(\alpha_{i j}\right)_{i, j=1}^s \Gamma^{-1}, \quad\left(c_{i j}\right)_{i, j=1}^s=\operatorname{diag}\left(\gamma_{11}^{-1}, \ldots, \gamma_{s s}^{-1}\right)-\Gamma^{-1},\\
	& \left(m_1, \ldots, m_s\right)=\left(b_1, \ldots, b_s\right) \Gamma^{-1}.
\end{align} 
\end{subequations}
This coefficient transformation \eqref{eA3} works in both directions (\textit{i.e.}, we can obtain either set of coefficients from the other). 
For the computational experiments in this article, we used the transformed ROS formulation listed in \eqref{eA5}. We list the coefficients for ROS3, ROW3, and \text{RO$\widetilde{W}$3} below.

\begin{align}\label{ROS3}
	&ROS3 \quad \text{\citep{SANDU19973459} }\notag \\
	&\begin{array}{|l|l|}
		\hline s=3 & m_1=1.0000000000000002\\
		\gamma=0.435866521508459& m_2= 6.1697947043828245\\
		& m_3= -0.42772256543218573 \\
		\hline a_{21}= 1.0 & c_{21}= -1.0156171083877703\\
		a_{31}=1.0 & c_{31}=4.07599564525377\\
		a_{32}= 0.0 & c_{32}=  9.20767942983308 \\\hline \alpha_{1}=0.0 & \gamma_{1}=\gamma\\
		\alpha_{2}=\gamma & \gamma_{2}=0.24291996454816805\\
		\alpha_{3}=\gamma & \gamma_{3}=2.185138002766406 \\
		\hline
	\end{array}
\end{align}
\begin{align}\label{ROS34PW1a}
	&\text{ROW3 (ROS34PW1a from \citep{rang_new_2005}})\notag\\
	&\begin{array}{|l|l|}
		\hline s=4 & m_1=6.1538321465310215  \\
		\gamma=0.435866521508459 & m_2=-0.8364233759732359 \\
		& m_3=-0.8614792120957679 \\
		& m_4=2.294280360279042 \\
		\hline a_{21}=5.0905205106702045 & c_{21}=-11.679081231228288    \\
		a_{31}=0.0 & c_{31}=-0.7100952636543062 \\
		a_{32}=0.0 & c_{32}=-0.04165460771675499\\
		a_{41}=  4.005173696367865 & c_{41}=-11.979557762226603 \\
		a_{42}= 0.19316470237944158  & c_{42}= -0.48054400523894975 \\
		a_{43}= 1.147140180139521   & c_{43}= 1.4350493021655284 
		\\\hline \alpha_{1}=0.0 & \gamma_{1}=\gamma\\
		\alpha_{2}=2.218787467653286& \gamma_{2}=-1.7829209461448272\\
		\alpha_{3}= 0.0& \gamma_{3}=0.33333333333333337 \\		\alpha_{4}=1.7837037931914073 & \gamma_{4}=-1.258070496147625 \\
		\hline
	\end{array}
\end{align}

  \begin{align}\label{GROW37n2}
 	&\text{RO$\widetilde{W}$3} \quad (\text{GROW37n2 from   \citep{jax2020rooted}})\notag\\
 	&\begin{array}{|l|l|}
 		\hline s=7 & m_1= 11.250177323676034\\
 		\gamma= 0.210537184485119 & m_2=3.192252948315025  \\
 		& m_3= 2.915007704973975\\
 		& m_4= 4.345658323293591 \\ & m_5= 3.012145125509720 \\& m_6=0.262455082551000\\
 		& m_7=1.0 \\
 		\hline a_{21}=1.999999999990039 & c_{21}=  -9.499509575398731   \\
 		a_{31}=4.019560642425581  & c_{31}= -12.665091304784792 \\
 		a_{32}= 1.345010761472820 & c_{32}=-4.246192603460309\\
 		a_{41}=   6.289460623523293 & c_{41}=  -11.670056440775877 \\
 		a_{42}= 2.085603163362063  & c_{42}= -4.621900088632271 \\
 		a_{43}=  0.819420955152500   & c_{43}= -0.786807019742143  \\
 		a_{51}=  7.190623951183970 & c_{51}=-10.908564086778629 \\
 		a_{52}= 1.331408706943976   & c_{52}= -3.828825206361854 \\
 		a_{53}=3.765300588143671    & c_{53}=  0.061767098865273
 		\\ 		a_{54}= -0.111788540780894   & c_{54}=  -8.852818969437871
 		\\
 		a_{61}=10.168196556048287 & c_{61}= -17.256895599015447 \\
 		a_{62}=3.269641580689367   & c_{62}= -5.337496593279378 \\
 		a_{63}=  1.330973976676946    & c_{63}= -4.213629752380526 \\ 
 		a_{64}= 3.873663222645613  & c_{64}= -16.380075552237702 \\
 		a_{65}=   1.585466384534608  & c_{65}= -11.765495488258829 \\  
 		a_{71}=  m_1 & c_{71}= -20.626172228646897 \\
 		a_{72}=m_2 & c_{72}=  -5.096509213224873 \\
 		a_{73}=m_3   & c_{73}=  -9.146293974318342\\  
 		a_{74}=  m_4 & c_{74}=-17.849863287382512 \\
 		a_{75}= m_5  & c_{75}= -16.208157949364576  \\
 		a_{76}= m_6  & c_{76}= -4.749754787719708 
 		\\\hline \alpha_{1}=0.0 & \gamma_{1}=\gamma\\
 		\alpha_{2}=   0.421074368968141
 		& \gamma_{2}=  -0.210537184483830
 		\\
 		\alpha_{3}= 0.563092201702538& \gamma_{3}=  -0.162638128403163
 		\\		\alpha_{4}=0.751799223121462 & \gamma_{4}=-0.074937378664796 \\\alpha_{5}=0.0 & \gamma_{5}=0.629578380496159\\
 		\alpha_{6}=1.0& \gamma_{6}= 0.0\\
 		\alpha_{7}= 1.0& \gamma_{7}=0.000000000000016\\ 
 		\hline
 	\end{array}
 \end{align}

\bibliographystyle{elsarticle-harv}
\bibliography{refs}

@article{Taylor1951,
        Author = {Taylor, GI},
        Journal = {Proc Roy Soc Lond Ser A},
        Pages = {447--461},
        Title = {Analysis of the swimming of microscopic organisms},
        Volume = 209,
        Year = 1951}

@article{Taylor1952,
        title = {The action of waving cylindrical tails in propelling microscopic organisms},
        volume = {211},
        issn = {0080-4630, 2053-9169},
        number = {1105},
        urldate = {2023-06-20},
        journal = {Proceedings of the Royal Society of London. Series A. Mathematical and Physical Sciences},
        author = {Taylor, Geoffrey Ingram},
        month = feb,
        year = {1952},
        pages = {225--239},
}

@article{Thomases_Guy_2017, title={The role of body flexibility in stroke enhancements for finite-length undulatory swimmers in viscoelastic fluids}, volume={825}, DOI={10.1017/jfm.2017.383}, journal={Journal of Fluid Mechanics}, author={Thomases, Becca and Guy, Robert D.}, year={2017}, pages={109–132}}

@article{Teran:2010,
        Author = {Teran, J. and Fauci, L. and Shelley, M.},
        Journal = {Phys. Rev. Lett.},
        Month = {Jan},
        Pages = {038101},
        Title = {Viscoelastic Fluid Response Can Increase the Speed and Efficiency of a Free Swimmer},
        Volume = {104},
        Year = {2010}}

@article{SANDU19973459,
title = {Benchmarking stiff ode solvers for atmospheric chemistry problems II: Rosenbrock solvers},
journal = {Atmospheric Environment},
volume = {31},
number = {20},
pages = {3459-3472},
year = {1997},
issn = {1352-2310},
doi = {https://doi.org/10.1016/S1352-2310(97)83212-8},
url = {https://www.sciencedirect.com/science/article/pii/S1352231097832128},
author = {A. Sandu and J.G. Verwer and J.G. Blom and E.J. Spee and G.R. Carmichael and F.A. Potra}
}

@article{ROSTAMI201660,
title = {Kernel-independent fast multipole method within the framework of regularized Stokeslets},
journal = {Journal of Fluids and Structures},
volume = {67},
pages = {60-84},
year = {2016},
issn = {0889-9746},
doi = {https://doi.org/10.1016/j.jfluidstructs.2016.07.006},
url = {https://www.sciencedirect.com/science/article/pii/S0889974615302450},
author = {Minghao W. Rostami and Sarah D. Olson}
}

@article{Walker_et_al_2019, title={Filament mechanics in a half-space via regularised Stokeslet segments}, volume={879}, DOI={10.1017/jfm.2019.723}, journal={Journal of Fluid Mechanics}, author={Walker, B. J. and Ishimoto, K. and Gadêlha, H. and Gaffney, E. A.}, year={2019}, pages={808–833}}

@book{hairer1993solving,
  title={Solving ordinary differential equations I: Nonstiff problems},
  author={Hairer, Ernst and Wanner, Gerhard and N{\o}rsett, Syvert P},
  year={1993},
  publisher={Springer},
  doi={https://doi.org/10.1007/978-3-540-78862-1}
}

@ARTICLE{Carichinoetal2021,
AUTHOR={Carichino, Lucia  and Drumm, Derek  and Olson, Sarah D. },
TITLE={A Computational Study of Hydrodynamic Interactions Between Pairs of Sperm With Planar and Quasi-Planar Beat Forms},
JOURNAL={Frontiers in Physics},
VOLUME={Volume 9 - 2021},
YEAR={2021},
URL={https://www.frontiersin.org/journals/physics/articles/10.3389/fphy.2021.735438},
DOI={10.3389/fphy.2021.735438},
ISSN={2296-424X}}

@article{ferranti2025,
title = {Analysis of the stability of an immersed elastic surface using the method of regularized Stokeslets},
journal = {Journal of Computational Physics},
volume = {550},
pages = {114636},
year = {2026},
issn = {0021-9991},
doi = {https://doi.org/10.1016/j.jcp.2025.114636},
url = {https://www.sciencedirect.com/science/article/pii/S0021999125009179},
author = {Dana Ferranti and Sarah D. Olson}
}

@article{rang_new_2005,
	title = {New {Rosenbrock} {W}-{Methods} of {Order} 3 for {Partial} {Differential} {Algebraic} {Equations} of {Index} 1},
	volume = {45},
	issn = {1572-9125},
	url = {https://doi.org/10.1007/s10543-005-0035-y},
	doi = {10.1007/s10543-005-0035-y},
	number = {4},
	journal = {BIT Numerical Mathematics},
	author = {Rang, J. and Angermann, L.},
	month = dec,
	year = {2005},
	pages = {761--787},
}

@misc{GIT_REP,
  author       = {Uddin, Moslem and Fauci, Lisa and Cortez, Ricardo and Buvoli, Tommaso},
  title        = {Improving-Performance-of-MRS-Simulations-Using-ROS-Methods},
  year         = {2026},
  publisher    = {GitHub},
  journal      = {GitHub repository},
  howpublished = {\url{https://github.com/muddin21/Improving-Performance-of-MRS-Simulations-Using-ROS-Methods}}}

@software{jax2018github,
  author = {James Bradbury and Roy Frostig and Peter Hawkins and Matthew James Johnson and Chris Leary and Dougal Maclaurin and George Necula and Adam Paszke and Jake Vander{P}las and Skye Wanderman-{M}ilne and Qiao Zhang},
  title = {{JAX}: composable transformations of {P}ython+{N}um{P}y programs},
  url = {http://github.com/jax-ml/jax},
  version = {0.3.13},
  year = {2018},
}

@article{Morshed25,
	title = {Filiform microswimmers in tubular confinements},
	author = {Morshed, Adnan and Cortez, Ricardo and Fauci, Lisa},
	journal = {Phys. Rev. Fluids},
	volume = {10},
	issue = {10},
	pages = {104903},
	numpages = {24},
	year = {2025},
	month = {Oct},
	publisher = {American Physical Society},
	doi = {10.1103/7t6v-w8sv},
	url = {https://link.aps.org/doi/10.1103/7t6v-w8sv}
}

@phdthesis{makanga23,
	TITLE = {{Transport and deformation of flexible fibers in structured environments}},
	AUTHOR = {Makanga, Ursy},
	URL = {https://theses.hal.science/tel-04543365},
	NUMBER = {2023IPPAX080},
	SCHOOL = {{Institut Polytechnique de Paris}},
	YEAR = {2023},
	MONTH = Sep,
	TYPE = {Theses},
	HAL_ID = {tel-04543365},
	HAL_VERSION = {v1},
}

@article{JABBARZADEH2020,
	title = {A numerical method for inextensible elastic filaments in viscous fluids},
	journal = {Journal of Computational Physics},
	volume = {418},
	pages = {109643},
	year = {2020},
	issn = {0021-9991},
	doi = {https://doi.org/10.1016/j.jcp.2020.109643},
	url = {https://www.sciencedirect.com/science/article/pii/S0021999120304174},
	author = {Mehdi Jabbarzadeh and Henry C. Fu}
}

@article{BRINGLEY20085397,
	title = {Validation of a simple method for representing spheres and slender bodies in an immersed boundary method for Stokes flow on an unbounded domain},
	journal = {Journal of Computational Physics},
	volume = {227},
	number = {11},
	pages = {5397-5425},
	year = {2008},
	issn = {0021-9991},
	doi = {https://doi.org/10.1016/j.jcp.2008.01.048},
	url = {https://www.sciencedirect.com/science/article/pii/S002199910800082X},
	author = {Thomas T. Bringley and Charles S. Peskin}}

@article{Teran09,
	author = {Teran, Joseph M. and Peskin, Charles S.},
	title = {Tether Force Constraints in Stokes Flow by the Immersed Boundary Method on a Periodic Domain},
	journal = {SIAM Journal on Scientific Computing},
	volume = {31},
	number = {5},
	pages = {3404-3416},
	year = {2009},
	doi = {10.1137/080720217},
	
	URL = { 
	
	https://doi.org/10.1137/080720217
	
	
	
	},
	eprint = { 
	
	https://doi.org/10.1137/080720217
	
	
	
	}
}

@article{Lighthill76,
	author = {Lighthill, James},
	title = {Flagellar Hydrodynamics},
	journal = {SIAM Review},
	volume = {18},
	number = {2},
	pages = {161-230},
	year = {1976},
	doi = {10.1137/1018040},
	
	URL = { 
	
	https://doi.org/10.1137/1018040
	
	
	
	},
	eprint = { 
	
	https://doi.org/10.1137/1018040
	
	
	
	}
}

@article{Nguyen01112014,
	author = {Hoa Nguyen and Ricardo Cortez and Lisa Fauci},
	title = {Computing Flows Around Microorganisms: Slender-Body Theory and Beyond},
	journal = {The American Mathematical Monthly},
	volume = {121},
	number = {9},
	pages = {810--823},
	year = {2014},
	publisher = {Taylor \& Francis},
	doi = {10.4169/amer.math.monthly.121.09.810},
	
	
	URL = { 
	
	https://doi.org/10.4169/amer.math.monthly.121.09.810
	
	
	
	},
	eprint = { 
	
	https://doi.org/10.4169/amer.math.monthly.121.09.810
	
	
	
	}
	
}

@article{ISHIKAWA_SIMMONDS_PEDLEY_2006, title={Hydrodynamic interaction of two swimming model micro-organisms}, volume={568}, DOI={10.1017/S0022112006002631}, journal={Journal of Fluid Mechanics}, author={ISHIKAWA, TAKUJI and SIMMONDS, M. P. and PEDLEY, T. J.}, year={2006}, pages={119–160}}

@article{MALHOTRA2024,
	title = {Efficient convergent boundary integral methods for slender bodies},
	journal = {Journal of Computational Physics},
	volume = {503},
	pages = {112855},
	year = {2024},
	issn = {0021-9991},
	doi = {https://doi.org/10.1016/j.jcp.2024.112855},
	url = {https://www.sciencedirect.com/science/article/pii/S0021999124001049},
	author = {Dhairya Malhotra and Alex Barnett}}

@book{Pozrikidis_1992, place={Cambridge}, series={Cambridge Texts in Applied Mathematics}, title={Boundary Integral and Singularity Methods for Linearized Viscous Flow}, publisher={Cambridge University Press}, author={Pozrikidis, C.}, year={1992}, collection={Cambridge Texts in Applied Mathematics}}

@article{Gray55,
	author = {Gray, J. and Hancock, G. J.},
	title = {The Propulsion of Sea-Urchin Spermatozoa},
	journal = {Journal of Experimental Biology},
	volume = {32},
	number = {4},
	pages = {802-814},
	year = {1955},
	month = {12},
	issn = {0022-0949},
	doi = {10.1242/jeb.32.4.802},
	url = {https://doi.org/10.1242/jeb.32.4.802},
}

@article{DORMAND198019,
 	title = {A family of embedded Runge-Kutta formulae},
 	journal = {Journal of Computational and Applied Mathematics},
 	volume = {6},
 	number = {1},
 	pages = {19-26},
 	year = {1980},
 	issn = {0377-0427},
 	doi = {https://doi.org/10.1016/0771-050X(80)90013-3},
 	url = {https://www.sciencedirect.com/science/article/pii/0771050X80900133},
 	author = {J.R. Dormand and P.J. Prince}
 }

@inbook{verwer1980rosenbrock,
	url = {https://doi.org/10.1515/9783486992977-013},
	title = {Rosenbrock methods and time-lagged Jacobian matrices},
	booktitle = {Band 11},
	author = {J. G. Verwer and S. Scholz},
	editor = {},
	publisher = {Oldenbourg Wissenschaftsverlag},
	address = {Berlin, Boston},
	pages = {173--184},
	doi = {doi:10.1515/9783486992977-013},
	isbn = {9783486992977},
	year = {1983},
	lastchecked = {2025-08-26}
}

@article{Atilim18,
	author  = {Atilim Gunes Baydin and Barak A. Pearlmutter and Alexey Andreyevich Radul and Jeffrey Mark Siskind},
	title   = {Automatic Differentiation in Machine Learning: a Survey},
	journal = {Journal of Machine Learning Research},
	year    = {2018},
	volume  = {18},
	number  = {153},
	pages   = {1--43},
	url     = {http://jmlr.org/papers/v18/17-468.html}
}

@phdthesis{jax2020rooted,
	title={A rooted-tree based derivation of ROW-type methods with non-exact Jacobian entries for index-one DAEs},
	author={Jax, Tim},
	year={2020},
	school={Dissertation, Wuppertal, Bergische Universit{\"a}t, 2019}
}

@article{steinebach_construction_2023,
	title = {Construction of Rosenbrock–Wanner method Rodas5P and numerical benchmarks within the Julia Differential Equations package},
	volume = {63},
	issn = {1572-9125},
	url = {https://doi.org/10.1007/s10543-023-00967-x},
	doi = {10.1007/s10543-023-00967-x},
	pages = {27},
	number = {2},
	journal = {{BIT} Numerical Mathematics},
	author = {Steinebach, Gerd},
	year = {2023},
}

@article{LIU2022111366,
	title = {Parallel-in-time simulation of biofluids},
	journal = {Journal of Computational Physics},
	volume = {464},
	pages = {111366},
	year = {2022},
	issn = {0021-9991},
	doi = {https://doi.org/10.1016/j.jcp.2022.111366},
	url = {https://www.sciencedirect.com/science/article/pii/S0021999122004284},
	author = {Weifan Liu and Minghao W. Rostami}
}

@article{Nourian_Shum_2023, title={A numerical method for the locomotion of bi-flagellated bacteria in viscous fluid}, volume={3}, DOI={10.1017/flo.2022.34}, journal={Flow}, author={Nourian, Vahid and Shum, Henry}, year={2023}, pages={E4}}

@article{BOUZARTH20104208,
	title = {A multirate time integrator for regularized Stokeslets},
	journal = {Journal of Computational Physics},
	volume = {229},
	number = {11},
	pages = {4208-4224},
	year = {2010},
	issn = {0021-9991},
	doi = {https://doi.org/10.1016/j.jcp.2010.02.006},
	url = {https://www.sciencedirect.com/science/article/pii/S0021999110000793},
	author = {Elizabeth L. Bouzarth and Michael L. Minion}
}

@article{Townsend_Keaveny_2024, title={Undulatory swimming in suspensions and networks of flexible filaments}, volume={993}, DOI={10.1017/jfm.2024.603}, journal={Journal of Fluid Mechanics}, author={Townsend, Adam K. and Keaveny, Eric E.}, year={2024}, pages={A10}}

@article{SCHOELLER2021,
	title = {Methods for suspensions of passive and active filaments},
	journal = {Journal of Computational Physics},
	volume = {424},
	pages = {109846},
	year = {2021},
	issn = {0021-9991},
	doi = {https://doi.org/10.1016/j.jcp.2020.109846},
	url = {https://www.sciencedirect.com/science/article/pii/S0021999120306203},
	author = {Simon F. Schoeller and Adam K. Townsend and Timothy A. Westwood and Eric E. Keaveny}
}

@inbook{butcherch3,
	author={J. C. Butcher},
	publisher = {John Wiley \& Sons, Ltd},
	isbn = {9781119121534},
	title = {Runge–Kutta Methods},
	booktitle = {Numerical Methods for Ordinary Differential Equations},
	chapter = {3},
	pages = {143-331},
	doi = {https://doi.org/10.1002/9781119121534.ch3},
	url = {https://onlinelibrary.wiley.com/doi/abs/10.1002/9781119121534.ch3},
	eprint = {https://onlinelibrary.wiley.com/doi/pdf/10.1002/9781119121534.ch3},
	year = {2016}
}

@article{Cortez01,
	author = {Cortez, Ricardo},
	title = {The Method of Regularized Stokeslets},
	journal = {SIAM Journal on Scientific Computing},
	volume = {23},
	number = {4},
	pages = {1204-1225},
	year = {2001},
	doi = {10.1137/S106482750038146X},
	
	URL = { 
	
	https://doi.org/10.1137/S106482750038146X
	
	
	
	},
	eprint = { 
	
	https://doi.org/10.1137/S106482750038146X
	
	
	
	}
}

@article{RANG2015128,
	title = {Improved traditional Rosenbrock–Wanner methods for stiff ODEs and DAEs},
	journal = {Journal of Computational and Applied Mathematics},
	volume = {286},
	pages = {128-144},
	year = {2015},
	issn = {0377-0427},
	doi = {https://doi.org/10.1016/j.cam.2015.03.010},
	url = {https://www.sciencedirect.com/science/article/pii/S0377042715001569},
	author = {Joachim Rang}
}

@article{ROS2,
	author = {Verwer, J.G. and Spee, E.J. and Blom, J. G. and Hundsdorfer, W.},
	title = {A Second-Order Rosenbrock Method Applied to Photochemical Dispersion Problems},
	journal = {SIAM Journal on Scientific Computing},
	volume = {20},
	number = {4},
	pages = {1456-1480},
	year = {1999},
	doi = {10.1137/S1064827597326651},
	
	URL = { 
	
	https://doi.org/10.1137/S1064827597326651
	
	
	
	},
	eprint = { 
	
	https://doi.org/10.1137/S1064827597326651
	
	
	
	}
}

@article{Cortez05,
	author = {Cortez, Ricardo and Fauci, Lisa and Medovikov, Alexei},
	title = "{The method of regularized Stokeslets in three dimensions: Analysis, validation, and application to helical swimming}",
	journal = {Physics of Fluids},
	volume = {17},
	number = {3},
	pages = {031504},
	year = {2005},
	month = {02},
	issn = {1070-6631},
	doi = {10.1063/1.1830486},
	url = {https://doi.org/10.1063/1.1830486},
	eprint = {https://pubs.aip.org/aip/pof/article-pdf/doi/10.1063/1.1830486/15742164/031504\_1\_online.pdf},
}

@Inbook{Lang2021,
	author="Lang, Jens",
	editor="Jax, Tim
	and Bartel, Andreas
	and Ehrhardt, Matthias
	and G{\"u}nther, Michael
	and Steinebach, Gerd",
	title="Rosenbrock-Wanner Methods: Construction and Mission",
	bookTitle="Rosenbrock---Wanner--Type Methods: Theory and Applications",
	year="2021",
	publisher="Springer International Publishing",
	address="Cham",
	pages="1--17",
	isbn="978-3-030-76810-2",
	doi="10.1007/978-3-030-76810-2_1",
	url="https://doi.org/10.1007/978-3-030-76810-2_1"
}

@article{CORTEZ201541,
	title = {A general system of images for regularized Stokeslets and other elements near a plane wall},
	journal = {Journal of Computational Physics},
	volume = {285},
	pages = {41-54},
	year = {2015},
	issn = {0021-9991},
	doi = {https://doi.org/10.1016/j.jcp.2015.01.019},
	url = {https://www.sciencedirect.com/science/article/pii/S0021999115000236},
	author = {Ricardo Cortez and Douglas Varela}
}

@book{wanner1996,
	title={Solving ordinary differential equations II},
	author={Ernst Hairer and Gerhard Wanner},
	volume={375},
	year={1996},
	publisher={Springer Berlin, Heidelberg},
	doi={https://doi.org/10.1007/978-3-642-05221-7}
}

@article{SIMONS201481,
	title = {The dynamics of sperm detachment from epithelium in a coupled fluid-biochemical model of hyperactivated motility},
	journal = {Journal of Theoretical Biology},
	volume = {354},
	pages = {81-94},
	year = {2014},
	issn = {0022-5193},
	doi = {https://doi.org/10.1016/j.jtbi.2014.03.024},
	url = {https://www.sciencedirect.com/science/article/pii/S0022519314001635},
	author = {Julie Simons and Sarah Olson and Ricardo Cortez and Lisa Fauci}
}

@article{OLSON2011203,
	title = {Coupling biochemistry and hydrodynamics captures hyperactivated sperm motility in a simple flagellar model},
	journal = {Journal of Theoretical Biology},
	volume = {283},
	number = {1},
	pages = {203-216},
	year = {2011},
	issn = {0022-5193},
	doi = {https://doi.org/10.1016/j.jtbi.2011.05.036},
	url = {https://www.sciencedirect.com/science/article/pii/S0022519311002888},
	author = {Sarah D. Olson and Susan S. Suarez and Lisa J. Fauci}
}

@article{FAUCI198885,
	title = {A computational model of aquatic animal locomotion},
	journal = {Journal of Computational Physics},
	volume = {77},
	number = {1},
	pages = {85-108},
	year = {1988},
	issn = {0021-9991},
	doi = {https://doi.org/10.1016/0021-9991(88)90158-1},
	url = {https://www.sciencedirect.com/science/article/pii/0021999188901581},
	author = {Lisa J Fauci and Charles S Peskin}
}

@article{Lang01,
	title = {{ROS3P}—{An} {Accurate} {Third}-{Order} {Rosenbrock} {Solver} {Designed} for {Parabolic} {Problems}},
	volume = {41},
	issn = {1572-9125},
	url = {https://doi.org/10.1023/A:1021900219772},
	doi = {10.1023/A:1021900219772},
	number = {4},
	journal = {BIT Numerical Mathematics},
	author = {Lang, J. and Verwer, J.},
	month = sep,
	year = {2001},
	pages = {731--738},
}

@Inbook{Wanner1977,
	author="Wanner, Gerhard",
	editor="Descloux, Jean
	and Marti, J{\"u}rg",
	title="On the integration of stiff differential equations",
	bookTitle="Numerical Analysis: Proceedings of the Colloquium on Numerical Analysis Lausanne, October 11--13, 1976",
	year="1977",
	publisher="Birkh{\"a}user Basel",
	address="Basel",
	pages="209--226",
	isbn="978-3-0348-5575-4",
	doi="10.1007/978-3-0348-5575-4_11",
	url="https://doi.org/10.1007/978-3-0348-5575-4_11"
}

@article{CORTEZ2018,
	title = {Regularized Stokeslet segments},
	journal = {Journal of Computational Physics},
	volume = {375},
	pages = {783-796},
	year = {2018},
	issn = {0021-9991},
	doi = {https://doi.org/10.1016/j.jcp.2018.08.055},
	url = {https://www.sciencedirect.com/science/article/pii/S0021999118305886},
	author = {Ricardo Cortez}
}

@article{Rosenbrock63,
	author = {Rosenbrock, H. H.},
	title = "{Some general implicit processes for the numerical solution of differential equations}",
	journal = {The Computer Journal},
	volume = {5},
	number = {4},
	pages = {329-330},
	year = {1963},
	month = {01},
	issn = {0010-4620},
	doi = {10.1093/comjnl/5.4.329},
	url = {https://doi.org/10.1093/comjnl/5.4.329},
	eprint = {https://academic.oup.com/comjnl/article-pdf/5/4/329/899502/050329.pdf},
}

@article{Keller_Rubinow_1976, title={Slender-body theory for slow viscous flow}, volume={75}, DOI={10.1017/S0022112076000475}, number={4}, journal={Journal of Fluid Mechanics}, author={Keller, Joseph B. and Rubinow, Sol I.}, year={1976}, pages={705–714}}

@article{yijiang-graham,
  title = {Free-space and near-wall dynamics of a flexible sheet sedimenting in Stokes flow},
  author = {Yu, Yijiang and Graham, Michael D.},
  journal = {Phys. Rev. Fluids},
  volume = {9},
  issue = {5},
  pages = {054104},
  numpages = {19},
  year = {2024},
  month = {May},
  publisher = {American Physical Society},
  doi = {10.1103/PhysRevFluids.9.054104},
  url = {https://link.aps.org/doi/10.1103/PhysRevFluids.9.054104}
}

@article{LAGRONE201973,
title = {Complex dynamics of long, flexible fibers in shear},
journal = {Journal of Non-Newtonian Fluid Mechanics},
volume = {269},
pages = {73-81},
year = {2019},
issn = {0377-0257},
doi = {https://doi.org/10.1016/j.jnnfm.2019.06.007},
url = {https://www.sciencedirect.com/science/article/pii/S0377025719301120},
author = {John LaGrone and Ricardo Cortez and Wen Yan and Lisa Fauci},
}

@article{Smith09,
    author = {Smith, D. J.},
    title = {A boundary element regularized Stokeslet method applied to cilia- and flagella-driven flow},
    journal = {Proceedings of the Royal Society A: Mathematical, Physical and Engineering Sciences},
    volume = {465},
    number = {2112},
    pages = {3605-3626},
    year = {2009},
    month = {09},
}

@article{gallagher2023,
  title = {Axonemal Regulation by Curvature Explains Sperm Flagellar Waveform Modulation},
  author = {Gallagher, Meurig T and Kirkman-Brown, Jackson C and Smith, David J},
  editor = {Amon, Cristina},
  date = {2023-03-03},
  journaltitle = {PNAS Nexus},
  volume = {2},
  number = {3},
  pages = {pgad072},
  issn = {2752-6542},
}

@article{GGSK2010,
    author = {Gadêlha, H. and Gaffney, E. A. and Smith, D. J. and Kirkman-Brown, J. C.},
    title = {Nonlinear instability in flagellar dynamics: a novel modulation mechanism in sperm migration?},
    journal = {Journal of The Royal Society Interface},
    volume = {7},
    number = {53},
    pages = {1689-1697},
    year = {2010},
    month = {05},
    issn = {1742-5689},
    doi = {10.1098/rsif.2010.0136},
    url = {https://doi.org/10.1098/rsif.2010.0136},
    eprint = {https://royalsocietypublishing.org/rsif/article-pdf/7/53/1689/496680/rsif.2010.0136.pdf},
}

@article{ratsperm,
    author = {Schmitz-Lesich, K.A. and Lindemann, C.B.},
    title = {Direct measurement of the passive stiffness of rat sperm and implications to the mechanism of the calcium response.},
    journal = {Cell Motil Cytoskeleton},
    volume = {59},
    number = {3},
    pages = {169-179},
    year = {2004},
}

@article{bullsperm,
    author = {Lesich, K.A. and Pelle, D. and  Lindemann, C.B.},
    title = {Insights into the mechanism of ADP
action of flagellar motility derived from studies of bull sperm.},
    journal = {Biophys. J.},
    volume = {95},
    pages = {472-482},
    year = {2008},
}

@article{humansperm,
    author = {Serres, C. and Escalier, D. and David, G.},
    title = {Ultrastructural morphometry of the human sperm flagellum with a stereological analysis of the lengths of the dense fibres.},
    journal = {Biol. Cell},
    volume = {49},
    number = {2},
    pages = {153-161},
    year = {1983},
}

@article{micesperm-radius,
    author = {Gervasi, María G. and Xu, Xinran and Carbajal-Gonzalez, Blanca and Buffone, Mariano G. and Visconti, Pablo E. and Krapf, Diego},
    title = {The actin cytoskeleton of the mouse sperm flagellum is organized in a helical structure},
    journal = {Journal of Cell Science},
    volume = {131},
    number = {11},
    pages = {jcs215897},
    year = {2018},
    month = {06},
    issn = {0021-9533},
    doi = {10.1242/jcs.215897},
    url = {https://doi.org/10.1242/jcs.215897},
    eprint = {https://journals.biologists.com/jcs/article-pdf/131/11/jcs215897/3502566/jcs215897.pdf},
}

\end{document}